\documentclass[12pt]{JHEP3}

\def\be{\begin{equation}}
\def\ee{\end{equation}}
\def\bea{\begin{eqnarray}}
\def\eea{\end{eqnarray}}
\def\be{\begin{equation}}
\def\ee{\end{equation}}
\def\nn{\nonumber}
\def\la{\langle}
\def\ra{\rangle}
\def\de{\partial}

\def\tr{{\rm tr}}
\def\sla{\raise.15ex\hbox{$/$}\kern-.57em}
\def\ie{{\it i.e.}~}
\def\eg{{\it e.g.}~}
\def\ap{{\alpha^\prime}}

\def\a{\alpha}
\def\b{\beta}

\def\d{\delta}
\def\e{\epsilon}

\def\et{\eta}
\def\th{\theta}

\def\l{\lambda}
\def\m{\mu}
\def\n{\nu}
\def\r{\rho}
\def\s{\sigma}
\def\t{\tau}

\def\ps{\psi}

\def\ve{\varepsilon}

\def\G{\Gamma}

\def\cA{{\cal A}}
\def\cB{{\cal B}}

\def\cE{{\cal E}}
\def\cF{{\cal F}}
\def\cG{{\cal G}}
\def\cH{{\cal H}}
\def\cI{{\cal I}}
\def\cJ{{\cal J}}

\def\cM{{\cal M}}
\def\cN{{\cal N}}

\def\cP{{\cal P}}

\def\cR{{\cal R}}
\def\cS{{\cal S}}
\def\cT{{\cal T}}
\def\cU{{\cal U}}
\def\cV{{\cal V}}
\def\cW{{\cal W}}
\def\cX{{\cal X}}

\def\cZ{{\cal Z}}

\author{Massimo Bianchi and Alessio V. Santini \\
Dipartimento di Fisica \& Sezione I.N.F.N.\\
Universit\`a di Roma
``Tor Vergata'' \\
Via della Ricerca Scientifica, 1 - 00133  Roma ITALY\\
E-mail: \email{Massimo.Bianchi@roma2.infn.it, avs980@katamail.com}}
 \abstract{We derive compact expressions for one-loop
scattering amplitudes of four open-string vector bosons around
supersymmetric configurations with intersecting or magnetized
D-branes on toroidal orbifolds. We check the validity of our
formulae against the structure of their singularities and their
behaviour under modular transformations to the transverse channel,
exposing closed string exchange. We then specialize to the case of
forward scattering and compute the total cross section for two
massless open string vector bosons on the brane to decay into
closed strings in the bulk, relying on the optical theorem.
Although not directly related to collider signatures our
predictions represent a step forward towards unveiling
phenomenological implications of open and unoriented
superstrings.} \preprint{ROM2F/2006/18
 \\ hep-th/0607224}
\title{String predictions for near future colliders from
one-loop scattering amplitudes around D-brane worlds}

\begin{document}

\section{Introduction and conclusions}

Vacuum configurations with open and unoriented strings have proven
to be a particularly rich arena where to address compelling
phenomenological issues in a string context amenable to explicit
computations (see \eg \cite{Dudas:2000bn, Angelantonj:2002ct, Uranga:2003pz,
Kiritsis:2003mc, Kokorelis:2004xm, Blumenhagen:2005mu} for
comprehensive reviews).

After their systematization \cite{Sagnotti:1987tw, Bianchi:1988fr,
Bianchi:1989du, Pradisi:1988xd, Bianchi:1990yu, Bianchi:1990tb,
Bianchi:1991rd, Bianchi:1991eu, mbphd}, including subtle effects
such as rank reduction induced by a quantized NS-NS antisymmetric
tensor \cite{Bianchi:1991eu, mbphd, Bianchi:1997rf, Witten:1997bs}
and the minimal coupling of RR p-form potentials in an asymmetric
superghost picture \cite{Bianchi:1991eu, mbphd}, these theories
have received an enormous boost when their geometric description
in terms of D-branes and $\Omega$-planes \cite{Polchinski:1995mt,
Gimon:1996rq}, pioneered in \cite{Dai:1989ua, Leigh:1989jq}, has
catalyzed the attention of the community.

Compactifications on toroidal orbifolds
\cite{Berkooz:1996dw,Angelantonj:1996uy, Kakushadze:1997ku,
Kakushadze:1997uj, Aldazabal:1998mr, Bianchi:1999uq} with
intersecting \cite{Berkooz:1996km, Cvetic:2001nr, Cvetic:2002pj,
Blumenhagen:2004vz, Angelantonj:2005hs} and/or magnetized
\cite{Balasubramanian:1996uc, Larosa:2003mz, Marchesano:2004yq,
Marchesano:2004xz, Dudas:2005jx} branes represent a simple yet
interesting class of models where problems connected with the
presence of (large) extra dimensions \cite{Antoniadis:1998ig,
Arkani-Hamed:1998nn, Dienes:1998vg}, supersymmetry breaking
\cite{Bachas:1995ik, Bianchi:1997gt, Antoniadis:1998ki,
Antoniadis:1998ep, Angelantonj:2004yt} and moduli stabilization
\cite{Giddings:2001yu, Kachru:2002he, Blumenhagen:2003vr,
Cascales:2003zp, Antoniadis:2004pp, Blumenhagen:2005tn,
Bianchi:2005yz, Derendinger:2005ph, Antoniadis:2005nu,
Camara:2005dc} can be tackled in a controllable way. The study of
interactions that determine the structure of the low energy
effective action and higher derivative corrections thereof has
lead to enormous effort at tree (disk and sphere) level
\cite{Klebanov:2003my, Cremades:2003qj, Cvetic:2003ch,
Cremades:2004wa, Lust:2004cx, Bertolini:2005qh} but relatively
little is known at one-loop \cite{Bachas:1996zt,
Antoniadis:1999ge, Lust:2003ky, Abel:2004ue, Bianchi:2005sa} and
beyond \cite{Berkovits:2006bk}. Aim of the present investigation
is to improve the situation in view of potential phenomenological
application of this kind of analyses in formulating predictions
for near future colliders based on models with relatively small
string tension and /or large extra dimensions \cite{Dudas:1999gz,
Cullen:2000ef, Accomando:1999sj, Chialva:2005gt}.

With this goal in mind, we use the standard NSR formalism
\cite{Friedan:1985ge} to derive compact expressions for one-loop
scattering amplitudes of four open-string vector bosons around vacuum
configurations with open and unoriented strings preserving some
supersymmetry. Remarkably, in addition to the standard elliptic
functions, our final formulae only involve two more modular forms
denoted by $\cE_\cN$ and $\cJ_\cN$ in the following. We check the
validity of our results, that generalize and extend the classic
results of Green and Schwarz in $D=10$ \cite{Green:1981ya,
Schwarz:1982jn}, against the structure of their singularities and
the properties under modular transformations to the transverse
channel that exposes closed string tree-level exchange.

We then compute the total cross section $\s_{tot} (s)$ for two
(massless) open string vector bosons on the brane to decay into
closed strings in the bulk around general unoriented vacuum
configurations preserving at least $\cN = 1$ supersymmetry. The
optical theorem relates $\s_{tot} (s)$ to the imaginary part of
the foward scattering amplitude $\cA (s)$. We work to lowest order
in $g_s$, \ie $\vert$disk$\vert^2\approx$ Im(annulus). In
principle, one could directly compute the amplitudes for the decay
of two massless open string states into (massive) closed string
states on the disk. These amplitudes are however plagued with
subtle normalization problems and we find it more convenient to
extract them from the non-planar one-loop forward scattering
amplitudes \cite{Dudas:1999gz, Cullen:2000ef, Accomando:1999sj,
Chialva:2005gt}.

The relevant contribution to the process corresponds to an
amplitude with vector bosons with the same Chan-Paton factors that
can annihilate into gauge-singlet closed string states. More
general initial states, involving massless open string scalars or
fermions can be computed similarly as described in a forthcoming
paper \cite{mbavsfermi}. Processes initiated by gauginos, in the
Adjoint of the Chan-Paton group, or matter scalars, in chiral
multiplets, are not relevant for collider physics. In addition to
the presently studied processes initiated by vector bosons,
transforming in the Adjoint of the Chan-Paton group and belonging
to the untwisted sector of the orbifold with integer modes
connecting parallel or equally magnetized branes, one should also
consider processes initiated by matter fermions, that may either
belong to the untwisted sector or to twisted sectors connecting
branes intersecting at angles or with different magnetic fluxes.
The latter can be studied in parallel and will be the focus of
\cite{mbavsfermi}.

Our present analysis exploits remarkable properties of elliptic
functions \cite{Eguchi:1986sb} and free field propagators on genus
one surfaces with or without boundaries and crosscaps to compute
and simplify the worldsheet correlators. After deriving compact
expressions for the one-loop non-planar amplitudes under
consideration, we specialize them to the case of forward
scattering. Extracting their imaginary part, we determine the
total cross section for the deacy into closed strings as a
function of the remaining Mandelstam variable $s = -(p_1 +p_2)^2 =
- 2 p_1 \cdot p_2$. It shows the expected peak and threshold
structure that encodes the properties of the brane configuration
in the internal space.

The plan of the paper is as follows. In Section \ref{treelev} we
recall basic formulae for vertex operators and tree level
scattering amplitudes. In Section \ref{oneloop} we compute the
one-loop contractions relevant for CP even processes. Section
\ref{summa} contains a summary of the results for the reader who
is not interested in the details of the derivations. CP odd
amplitudes receiving contribution only from $\cN =1$ sectors are
discussed in section \ref{CPodd}. The case of forward scattering
both for CP even and CP odd amplitudes is studied in section
\ref{forwscatt}, while the total cross sections for the decay into
closed strings in the bulk are discussed in section
\ref{totcross}. We present our final comments and draw
perspectives for our future investigation in section \ref{comm}.

Some properties of elliptic functions are collected in the
appendices that should be consulted for notation and conventions.

\section{Tree level amplitude}
\label{treelev}

For completeness, comparison and later purposes we report here the
results for the tree level (disk) scattering amplitude of four
open string gauge bosons. The result is independent of the amount
of supersymmetry enjoyed by the vacuum configuration, since the
relevant vertex operators in the NSR formalism\footnote{Unless
otherwise stated, we set $\ap = 1/2$ henceforth.}
 \bea &&V^{a\tilde{a}}_{0}(z) = a^\m (
\de X_\m +i p \cdot \ps \ps_\m )
\ e^{ip\cdot X} (z) T^{a\tilde{a}}_{CP}\ , \nn \\
&&V^{a\tilde{a}}_{-1}(z) = a^\m \ps_\m e^{-\varphi} \ e^{ip\cdot
X} (z) T^{a\tilde{a}}_{CP} \ , \eea with $p^2=0$, $a\cdot p =0$
for BRS invariance and $\tilde{a}=\Omega a$, do not depend on the
details of the compactification encoded in the internal CFT, to be
specified later on. In other words they only involve the identity
operator that has trivial correlators. The matrices
$T^{a\tilde{a}}_{CP}$ belong to the adjoint of the Chan-Paton
group, \ie to the ${\bf N}\bar{\bf N}$ for $U(N)$ or to the ${\bf
N(N\pm 1)/2}$ for $Sp(N)$ or $SO(N)$.

Vertex operators are inserted on the boundary of the disk that is
conformally equivalent to the upper half plane whereby $z_i = x_i
\in {\bf R}$. Three $c$ ghost insertions are needed to fix
$SL(2,R)$ invariance. The tree level four bosons scattering
amplitude is then given by \bea &&\cA^{tree}_{vv\rightarrow vv}
(p_i,a_i)= \\&& g_s \tr(T_1 T_2 T_3 T_4) \int dz_3 \la c
V_0(z_1;p_1,a_1) c V_{-1}(z_2; p_2,a_2) V_0(z_3; p_3,a_3) c
V_{-1}(z_4; p_4,a_4)\ra \ , \nn \eea up to permutations of the
external legs.

After performing the free field contractions and including the
relevant non-cyclic permutations, one finds\footnote{We use mostly
plus signature for the Lorentz metric $\eta_{\mu\nu}=(-,+,+,+)$.}
\bea &&\cA^{tree}_{vv\rightarrow vv} (p_i,a_i) = g_s (2\pi)^4
\d(\sum_i p_i) K^{tree}_{vv\rightarrow vv}(p_i,a_i)
\times \nn \\
&&\{[\tr(T_1 T_2 T_3 T_4) + \tr(T_1 T_4 T_3 T_2)] B(s,t) + \nn
\\
&& [\tr(T_1 T_3 T_4 T_2) + \tr(T_1 T_2
T_4 T_3)] B(t,u) + \nn \\
&& [\tr(T_1 T_4 T_2 T_3)+ \tr(T_1 T_3 T_2 T_4)]  B(u,s)\} \ ,\eea
where \be B(s,t) = \int_0^1 dx x^{2\ap p_1\cdot p_2-1} (1-x)^{2\ap
p_1\cdot p_3-1} = {\G(-\ap s) \G(-\ap t) \over \G (\ap u)} \ee is
Euler Beta function that appears in the celebrated Veneziano
amplitude, $s = - (p_1+p_2)^2 = -2 p_1\cdot p_2$, $t = -
(p_1+p_4)^2 = -2 p_1\cdot p_4$, $u = - (p_1+p_3)^2 = -2 p_1\cdot
p_3$ and thus $s+t+u = 0$.

The bosonic kinematic factor $K^{tree}_{vv\rightarrow
vv}(p_i,a_i)$ is totally symmetric (not simply cyclically
symmetric!) and reads \cite{Green:1981ya, Schwarz:1982jn} \bea &&
K^{tree}_{vv\rightarrow vv}(p_i,a_i) = -{1\over 4} ( st a_1\cdot
a_3 a_2\cdot a_4 + us a_1\cdot a_4
a_2\cdot a_3 + tu a_1\cdot a_2 a_3\cdot a_4 ) \\
&&+{s\over 2}  ( a_1\cdot p_4 a_3\cdot p_2 a_2\cdot a_4 + a_2\cdot
p_3 a_4\cdot p_1 a_1\cdot a_3 + a_1\cdot p_3 a_4\cdot p_2 a_2\cdot
a_3 + a_2\cdot p_4 a_3\cdot p_1 a_1\cdot a_4) \nn \\
&&+{t\over 2}  ( a_2\cdot p_1 a_4\cdot p_3 a_3\cdot a_1 + a_3\cdot
p_4 a_1\cdot p_2 a_2\cdot a_4 + a_2\cdot p_4 a_1\cdot p_3 a_3\cdot
a_4 + a_3\cdot p_1 a_4\cdot p_2 a_2\cdot a_1) \nn \\
&&+{u\over 2}  ( a_1\cdot p_2 a_4\cdot p_3 a_3\cdot a_2 + a_3\cdot
p_4 a_2\cdot p_1 a_1\cdot a_4 + a_1\cdot p_4 a_2\cdot p_3 a_3\cdot
a_4 + a_3\cdot p_2 a_4\cdot p_1 a_2\cdot a_1) \nn  \eea in
dimension $D = 10$ as well as in lower dimensions. It may be
written more compactly in terms of the linearized field strengths
\be f^{\mu\nu}_i = p^\mu_i a^\nu_i - p^\nu_i a^\mu_i = -
f^{\nu\mu}_i\ee thus getting a manifestly gauge invariant
expression \bea &&K^{tree}_{vv\rightarrow vv}(a_i,p_i) = {1\over
2} [(f_1 f_2 f_3 f_4) +
(f_1 f_3 f_4 f_2) + (f_1 f_4 f_2 f_3)] \nn \\
&&- {1\over 4}[(f_1 f_2) (f_3 f_4) + (f_1 f_3) (f_4 f_2) + (f_1
f_4)( f_2 f_3)] \eea where \be (f_i f_j f_k f_l)= f_i^\mu{}_\nu
f_j^\nu{}_\rho f_k^\rho{}_\sigma f_l^\sigma{}_\mu \ee and \be (f_i
f_j)= f_i^\mu{}_\nu f_j^\nu{}_\mu \ .\ee

 For forward scattering $p_4 = -p_1$ and $p_3 = -p_2$,
so that $t=0$ and $u=-s$, $a_3=a_2$ and $a_4=a_1$, the kinematic
factor drastically simplifies to \be K^{FS,tree}_{vv\rightarrow
vv} = (\ap s)^2 a_1^2 a_2^2 \ . \ee

\section{One-loop CP even amplitudes}
\label{oneloop}

 In this section we compute the worldsheet correlators that appear
in the one-loop scattering amplitudes (planar, non-planar and
non-orientable) of four open string vector bosons for
supersymmetric models with intersecting and magnetized branes on
orbifolds\footnote{Some properties of elliptic functions and
propagators are collected in the Appendix that should be consulted
for notation and conventions.}. For the moment we focus on CP even
amplitudes that recieve contribution from the even spin structure
in the NSR formalism. CP odd amplitudes from the odd spin
structure are described in the next section.

Up to Chan-Paton factors, the one-loop four vector boson amplitude
in the direct channel (`open string' description) reads

\bea &&\cA^{1-loop}_{vv\rightarrow vv} (p_i,a_i) = g_s^2
\int_0^\infty {dt \over t} \int_{\cR} \prod_i dz_i  \nn
\\&&\sum_\a c_\a \la V_0(z_1; p_1,a_1) V_0(z_2; p_2,a_2)V_0(z_3;
p_3,a_3)V_0(z_4; p_4,a_4)\ra_\a \ . \eea

The power of the modular parameter $t$ in the denominator takes
care of the volume of the conformal Killing group and effectively
cancels the integration over the `center of mass' coordinate the
correlator is independent of. Additional negative powers of $t$
will appear as a result of integration over loop momentum.
Summation over the spin structures $\a$ ($\a =2,3,4$ even, $\a=1$
odd) with appropriate coefficients $c_\a$ implements the GSO
projection. Moreover summation over the various kinds of
magnetized or intersecting branes, later on labelled by $a= 1,...,
N_a$ with $N_a = Tr_a({\bf 1})$, and the various sectors of the
orbifold, later on labelled by $k=0, ...,n-1$ for the case of
$\Gamma = {\bf Z}_n$, is understood.

In the planar case, all the four vectors should belong to the same
factor in the Chan-Paton group and the annulus amplitude $\cA$ is
schematically given by \be \cA^{plan}_{vv\rightarrow vv} =
\tr_a(T_1 T_2 T_3 T_4 W_k) \tr_b(W_k) \cA_{ab}(1,2,3,4) \ , \ee
where the discrete Wilson lines $W$ represent the projective
embedding of the orbifold group $\Gamma$ in the Chan-Paton group.
The open string vertices are inserted on the same boundary of the
worldsheet, $z= -\bar{z}$, and the integration region is given by
\be \cR^{plan}_{\cA} = \{ z_i=iy_i :
0<y_1<y_2<y_3<y_4<Im\tau_\cA=t/2\} \ . \ee

Non-planar amplitudes receive contribution also when the vector
bosons belong to different factors of the Chan-Paton group, \ie
end on different stacks of D-branes. Up to permutations, depending
on the choice of the Chan-Paton matrices for the external legs,
the corresponding annulus amplitude $\cA$ reads \be
\cA^{nonpl}_{vv\rightarrow vv} = \tr_a(T_1 T_2) \tr_b(T_3 T_4)
\cA_{ab} (1,2;3,4) \ . \ee The integration region is given by is
\be \cR^{nonpl}_{\cA} = \{ z_{1,2}=iy_{1,2}, z_{3,4}={1\over 2} +
iy_{3,4} : 0<y_i<Im\tau_\cA=t/2 \} \ee but otherwise unrestricted.
This will play a crucial role later on in section \ref{totcross}.

For unoriented strings, the only ones where tadpole cancellation
can be achieved thanks to the contribution of the $\Omega$-planes,
one has to take into account the contribution of the M\"obius
strip $\cM$, too. Up to permutations, it reads \be
\cA^{unor}_{vv\rightarrow vv} = \tr_a(T_1 T_2 T_3 T_4
W^{\Omega}_{2k})\cM_{a\tilde{a}}(1,2,3,4) \ , \ee where
$W^{\Omega}_{2k}$ implements the action of the worldsheet parity
in the Chan-Paton group. The choice of $N$'s and $W$'s as well as
of the fluxes and intersection angles is tightly constrained by
consistency conditions such as RR-tadpole cancellation
\cite{Sagnotti:1992vb, Morales:1998ux,
Scrucca:1999uz,Aldazabal:1999nu, Bianchi:2000de}. We will assume
that such a choice has been made and shall not discuss this issue
any further in this paper. Moreover we will not consider phenomena
associated to the presence of anomalous $U(1)$'s discussed in
\cite{Ibanez:1998qp, Antoniadis:2002cs, Anastasopoulos:2003aj,
Anastasopoulos:2006cz}.

For the M\"obius strip $\cM$, the integration region is given by
\be {\cR}_{\cM} = \{ z_i=iy_i+{1\over 2} \delta: 0
<y_1<y_2<y_3<y_4<Im\tau_\cM = t/2; \delta =1,2 \} \ .\ee

Given the form of the vector emission vertices there are in
principle five different kinds of contributions to the worldsheet
correlators: \be \la (\de X)^4\ra_\a + ``4" \la (\de
X)^3(\ps\ps)\ra_\a + ``6"\la (\de X)^2 (\ps\ps)^2 \ra_\a + ``4"
\la (\de X)(\ps\ps)^3 \ra_\a + \la (\ps\ps)^4 \ra_\a
\label{correlators} \ . \ee It is easy to check that the first two
kinds of terms vanish for any supersymmetric vacuum configuration
after summation over the even spin structures or lack of fermionic
zero-modes in the odd spin structure. We are thus left with the
last three structures.

Contractions of the spacetime bosonic coordinates, satisfying
Neumann boundary conditions, are performed by means of \be
\cG_\Sigma(z-w)= {1\over 2}[ \cG_\cT(z-w) + \cG_\cT(z-\tilde{w}) +
\cG_\cT(\tilde{z}-w) + \cG_\cT(\tilde{z}-\tilde{w})]
\label{boseprop} \ , \ee where $\tilde{z} = 1 - \bar{z} =z$ and
$\tilde{w} = 1 - \bar{w} =w$ for open string insertions on the
boundary of $\Sigma= \cA, \cM$, and $\cG_\cT(z,w)$ is the bosonic
propagator (Bargmann kernel) on the covering torus
 \be
\cG_\cT(z-w) = -{\ap\over 2} \left[ \log\left\vert {\th_1(z) \over
\th_1'(0)}\right\vert^2 - {2\pi \over Im\tau} Im(z-w)^2 \right]
\label{boseproptor} \ , \ee with $\tau=\tau_\Sigma$.

In the even spin structures, free fermion contractions are
performed by means of \be \cS_\a(z-w) = {\th_\a(z-w) \over
\th_1(z-w)} { \th_1'(0) \over \th_\a(0)} \label{fermipropeven} \ ,
\ee the fermionic propagator (Szego kernel).

In the odd spin structure, the fermionic propagator may be taken
to be \be \cS_1 (z-w) = -\de_z \cG(z-w) \label{fermipropodd} \ .
\ee

Contractions are weighted by the partition function \be \cZ^\cN_\a
= \la 1 \ra^{a,b}_{\a,k} \ , \ee whose explicit form, as we will
momentarily see depends on the sector under consideration, \ie on
the choice of $k$ and $a, b$ that determine the number of
preserved supersymmetries $\cN$.

\subsection{ $\cN = 4$ sectors: only CP even amplitudes as in $D=10$}

Actually for ${\cN}= 4$ sectors only the last term in
(\ref{correlators}) contributes, \ie survives summation over the
(even) spin structures. The fermionic contractions give a constant
since the lowest derivative, $\ap \rightarrow 0$ limit,
four-vector amplitude is BPS saturated in these sectors.

${\cN}= 4$ open string sectors are characterized by $k=0$ and
connect parallel and equally magnetized branes (`neutral' and
`dipole' strings \cite{Angelantonj:2002ct, Bianchi:2005sa}). The
partition function is given by \be \cZ_\a^{{\cN}= 4} = \cX^{{\cN}=
4}_{ab}
 {\theta_\a^4(0) \over \eta^{12}} \ , \ee where
\be \cX^{{\cN}= 4}_{ab} = {\int d^4x_0 \Lambda_{ab} \over 2_{GSO}
2_\Omega n_{orb} (\ap t)^2} \label{XfactN4} \ee takes care of
numerical factors and bosonic zero modes. In particular, $\int
d^4x_0 = V_X$ is the (regulated) volume of spacetime, $\int d^4p_0
\exp(-\pi\ap p_0^2)= 1/(\ap t)^2$, and  $\Lambda_{ab}$ denotes the
6-dimensional sum over generalized KK momenta. The numerical
factors result from the various projections ${\bf Z}^{GSO}_2$ ,
${\bf Z}^\Omega_2$ and  ${\bf Z}^{orbifold}_n$.

Up to permutations, the (non)planar annulus amplitude reads

 \bea &&\cA_{ab}(1,2,3,4) (p_i,a_i) = g^2_s {(2\pi)^4 \over 4n}
\d(\sum_i p_i) K^{tree}_{vv\rightarrow vv}(p_i;a_i) \times \nn \\
&&\int_0^\infty {dt\over t^3} \Lambda_{ab}(\tau_\cA)
\int_{\cR^{(non)plan}_\cA}\prod_k dz_k \prod_{i<j} e^{-p_i\cdot
p_j \cG_\cA(z_{ij})} \ , \eea where $\cG_\cA$ is the free bosonic
propagator on the boundary of the annulus $\cA$ \ref{boseprop}.
The integration regions $\cR^{plan}_\cA$ and $\cR^{non-plan}_\cA$
have been discussed above. As indicated, the kinematic factor
$K^{tree}_{vv\rightarrow vv}$ is exactly the same as at tree
level.

The unoriented M\"obius amplitude reads
\bea &&\cM_{a\tilde{a}}(1,2,3,4) = g^2_s {(2\pi)^4
\over 4n}
\d(\sum_i p_i) K^{tree}_{vv\rightarrow vv}(p_i;a_i) \times \nn \\
&& \int_0^\infty {dt\over t^3} \Lambda_{{a\tilde{a}}}(\tau_\cA)
\int_{\tilde{\cR}_{\cM}}\prod_k dz_k \prod_{i<j} e^{-p_i\cdot p_j
\cG_\cM(z_{ij})} \ , \eea where $\cG_\cM$ is the free boson
propagator on the boundary of the M\"obius strip \ref{boseprop}.

\subsection{$\cN = 1,2$ sectors: CP even amplitudes}

Let us consider the $\cN=1$ and $\cN=2$ supersymmetric sectors
that can be analyzed in parallel in the even spin structures. The
odd spin structure needs a separate analysis. The main difference
between the two cases resides in the internal contribution . Our
analyses applies to arbitrary choices (`parallel'
\cite{Angelantonj:2002ct} or `oblique'  \cite{Bianchi:2005sa}) of
constant abelian magnetic fluxes\footnote{A worldsheet description
of the non abelian fluxes discussed in \cite{Kumar:2006er} is not
yet available.} or intersecting angles in (supersymmetric)
orbifold compactifications.

In the ${\cN}=1$ case one has \be \la 1 \ra_{\a, k, \ve}^{a,b} =
\cZ^{{\cN}=1}_\a(u^I_{ab}) = \cX^{{\cN}=1}_{ab} {\th_\a(0)\over
\eta^3}
 \prod_I {\th_\a(u^I_{ab})\over \th_1(u^I_{ab})} \ , \ee
where \be\cX^{{\cN}=1}_{ab} =   { \cI_{ab} \int d^4x_0 \over
2_{GSO} 2_\Omega n_{orb} (\ap t)^2} \, \label{XfactN1} \ee
$\cI_{ab}$ is some discrete multiplicity, \eg degeneracy of Landau
levels, number of fixed points or intersections, \be u^I_{ab} =
\ve^I_{ab} \tau_\cA + kv^I_{ab} \ , \ee with $\ve^I_{ab}$ denoting
intersection angles or magnetic shifts and $kv^I_{ab}$
implementing some ${\bf Z}_n$ orbifold projection with
$k=0,1,...n-1$. ${\cN}=1$ supersymmetry requires $u^I_{ab} \neq
0$, with $\sum_I u^I_{ab} = 0$ (mod 1).

In the $\cN=2$ case one of the $u^I_{ab}$ vanishes. For
definiteness let us set $u^3_{ab}= 0$, then $u^1_{ab}= - u^2_{ab}
= u_{ab}$  (mod 1) and one obtains \be \la 1 \ra_{\a, k,
\ve}^{a,b} = \cZ^{{\cN}=2}_\a(u_{ab}) = \cX^{{\cN}=2}_{ab}
{\th_\a(0)^2\th_\a(u_{ab})\th_\a(-u_{ab})\over \eta^6
\th_1(u_{ab})\th_1(-u_{ab})} \ , \ee where \be \cX^{{\cN}=2}_{ab}
= { \int d^4x_0 \cI_{ab}^\perp \Lambda^\parallel_{ab}(\tau_{\cA})
\over 2_{GSO} 2_\Omega n_{orb} (\ap t)^2}  \label{XfactN2} \ . \ee
In addition to the discrete multiplicity $\cI_{ab}^\perp$ in the
twisted or magnetized directions, a sum of generalized KK momenta
$\Lambda_{ab}^\parallel(\tau_{\cA})$ in the untwisted or
unmagnetized directions is present.

In both cases the Chan-Paton factors get modified to $
\tr_a(T^1...W^k)$ by the effect of (discrete) Wilson lines $W$
corresponding to the projective embedding of the orbifold group
$\Gamma$ in the gauge group. Notice that the unbroken gauge group
$G^u_a$ for branes of type $a$ corresponds to the generators
$T^u_a$ such that $[T^u_a,W_a^k]=0$.

Let us now consider the Wick contractions one at a time.

\subsubsection{Two fermion bilinears (6 terms)}

Up to permutations (six in all) the typical $\la (\de
X)^2(\ps\ps)^2 \ra_\a$ correlator reads \be \la a_1\cdot \de X
e^{ip_1\cdot X} (z_1) a_2\cdot \de X e^{ip_2\cdot X} (z_2)
e^{ip_3\cdot X} (z_3)e^{ip_4\cdot X}(z_4)\ra \la {ip_3\cdot
\ps}a_3\cdot\psi (z_3){ip_4\cdot \ps}a_4 \cdot \ps(z_4)\ra_\a \ .
\ee

The bosonic correlator yields  \bea && \la a_1\cdot \de X
e^{ip_1\cdot X} (z_1) a_2\cdot \de X e^{ip_2\cdot X} (z_2)
e^{ip_3\cdot X} (z_3) e^{ip_4\cdot X}(z_4) \ra = \nn \\&&[ a_1
\cdot a_2\de_1 \de_2 \cG(z_{12}) - \sum_{i\neq 1} a_1 \cdot p_i
\de_1 \cG(z_{1i})\sum_{j\neq 2} a_1 \cdot p_j \de_2 \cG(z_{2j})]
\Pi(p_i,z_i) \ , \eea where $\cG$ is the free bosonic propagator
defined in (\ref{boseprop}) and \be \Pi(p_i,z_i) = \prod_{i<j}
\exp [-p_i\cdot p_j \cG(z_{ij})] \label{momfact}\ .  \ee

 The
fermionic correlator yields \be \la {ip_3\cdot \ps}a_3\cdot\psi
(z_3){ip_4\cdot \ps}a_4 \cdot \ps(z_4)\ra_\a  = - 2 {1 \over 2^2}
(f_3 f_4) \cS^2_\a(z_{34}) \cZ_\a \ , \ee where $\cS_\a$ is the
fermionic propagator (Szego kernel) for even spin structures
defined in (\ref{fermipropeven}). Using \be \cS^2_\a(z-w)
=\cP(z-w) - e_{\a-1} \label{SStoP} \ , \ee where $\cP(z)$ is
Weierstrass $\cP$ function \be \cP(z) = - \de_z^2 \log\theta_1(z)
- 2\et_1 \ , \ee with \be \et_1 = -2\pi i \de_\tau \log\et = -
{1\over 6} {\theta_1'''(0) \over \theta_1'(0)} \label{eta1} \ ,
\ee since $\theta_1'(0) = 2\pi \eta^3$, and \be e_{\a-1} = -4\pi i
{d \over d\tau} \log{\theta_\a(0|\tau) \over \eta(\tau)} \ , \ee
it is easy to see that only the term $\de_\t\log\th_{\a}$ in
$e_{\a-1}$ survives summation over the even spin structures.

For ${\cN}=1$ sectors, one finds \be \cE_{{\cN}=1}(u^I_{ab}) =
\cX^{{\cN}=1}_{ab} \sum_\a {\th''_{\a}(0)\over \et^3}\prod_I
{\th_\a(u^I_{ab})\over \th_1(u^I_{ab})} = 2\pi \cX^{{\cN}=1}_{ab}
\sum_I {\th'_1(u^I_{ab})\over \th_1(u^I_{ab})} = 2\pi
\cX^{{\cN}=1}_{ab} {\cH'(0)\over \cH(0)} \label{Eterm1} \ , \ee
where \be \cH(z) = \prod_I \th_1(z + u^I_{ab})\label{hfunct} \ee
and the zero-mode factor $\cX_{ab}^{\cN = 1}$ is defined in
(\ref{XfactN1}).

For ${\cN}=2$ sectors one finds \be \cE_{{\cN}=2}(u_{ab}) =
\cX^{\cN =2}_{ab}\sum_\a
{\th''_{\a}(0)\th_{\a}(0)\th_\a(u_{ab})\th_\a(-u_{ab})\over \eta^3
\th_1(u_{ab})\th_1(-u_{ab})} = 4\pi^2 \cX^{\cN =2}_{ab}
\label{Eterm2} \ , \ee where the zero-mode factor $\cX_{ab}^{\cN =
2}$ is defined in (\ref{XfactN2}).

Thus, eventually the fermionic correlator  \be \la {ip_3\cdot
\ps}a_3\cdot\psi(z_3) {ip_4\cdot \ps}a_4\cdot \ps (z_4)\ra_{even}
= - {1 \over 2} (f_3 f_4)\cE_{{\cN}=1,2}(u^I_{ab}) \ee turns out
to be independent of the insertion points in ${\cN}=1,2$ sectors.
As already stated, it receives vanishing contribution from
${\cN}=4$ sectors.

\subsubsection{Three fermion bilinears (4 terms)}

Up to permutations (four in all) the correlator $\la (\ps\ps)^3
(\de X) \ra_\a$
 can be most
conveniently computed by observing that normal ordering of the
fermion bilinears $:\ps\ps:$ allows only for a cyclic Lorentz
contraction that yields \be \la {ip_1\cdot \ps}a_1\cdot\psi
{ip_2\cdot \ps}a_2\cdot \ps {ip_3\cdot \ps}a_3\cdot \ps \ra_\a = -
i {2^3 \over 2^3} (f_1f_2f_3)
\cS_\a(z_{12})\cS_\a(z_{23})\cS_\a(z_{13})\cZ^\cN_\a \ , \ee where
$z_{ij} = z_i - z_j$ and \be (f_1 f_2 f_3) =
f_1^{\m}{}_{\n}f_2^{\n}{}_{\r}f_3^{\r}{}_{\m} \ . \ee Using the
identity \be \cS_\a(z_{13})\cS_\a(z_{23}) = \cS_\a(z_{12})
\omega(z_1,z_2,z_3) + \cS'_\a(z_{12}) \label{SSident} \ , \ee
where \be \omega(z_1,z_2,z_3) = \de_1 \log\th_1(z_{12}) + \de_2
\log\th_1(z_{23}) + \de_3 \log\th_1(z_{31}) \ , \ee re-combining
the two $\cS_\a(z_{12})$, using (\ref{SStoP})
 and \be
\cS_\a(z)\cS'_\a(z) = {1\over 2}
\partial_z (\cP(z) - e_{\a -1}) = {1\over 2} \cP'(z) \ , \ee  and
summing over spin structures yields \be \la {ip_1\cdot
\ps}a_1\cdot\psi {ip_2\cdot \ps}a_2\cdot \ps {ip_3\cdot
\ps}a_3\cdot \ps \ra_{even} = - i (f_1 f_2 f_3)
\omega(z_1,z_2,z_3) \cE_{{\cN}}(u^I_{ab}) \ , \ee which is
manifestly symmetric under any permutation of the three insertion
points. As already stated, this correlator gets no contribution
from $\cN =4$ sectors.

The bosonic correlator simply yields \be \la e^{ip_1\cdot X}
e^{ip_2\cdot X} e^{ip_3\cdot X} a_4\cdot \de X e^{ip_4\cdot X}\ra
= i \sum_{i\neq 4} a_4\cdot p_i \de_4 \cG(z_{i4}) \Pi(p_i,z_i) \ ,
\ee where $\cG$ is the bosonic propagator \ref{boseprop} and
$\Pi(p_i,z_i)$ is the momentum factor (\ref{momfact}).

\subsubsection{Four fermion bilinears (1 term, 2 structures)}

Let us finally consider the $\la (\ps\ps)^4 \ra_\a$ term. The
bosonic coordinates contribute the momentum factor $\Pi(z_i;p_i)$
defined in (\ref{momfact}). Taking into account normal ordering of
the $:\ps\ps:$'s allows for two kinds of contraction.

Three are connected contractions of the Lorentz indices that yield
\be \la (\ps\ps)^4\ra_\a^{conn} = { 2^4\over 2^4} (f_1 f_2 f_3
f_4) \cS_\a(z_{12})\cS_\a(z_{23})\cS_\a(z_{34})
\cS_\a(z_{14})\cZ_\a \ , \ee where \be (f_1 f_2 f_3 f_4) =
f_1^{\m}{}_{\n}f_2^{\n}{}_{\r}f_3^{\r}{}_{\l}f_4^{\l}{}_{\m} \ .
\ee Using (\ref{SSident}), the product of fermionic propagators
can be simplified to \bea &&[\cS_\a(z_{13})\omega(z_1,z_2,z_3) +
\cS'_\a(z_{13})]
[\cS_\a(z_{13})\omega(z_1,z_4,z_3) + \cS'_\a(z_{13})] \nn\\
= && \omega(z_1,z_2,z_3)\omega(z_1,z_4,z_3)[\cP(z_{13}) - e_{\a
-1}
]  \\
&& +   {1\over 2} [\omega(z_1,z_2,z_3) +\omega(z_1,z_4,z_3)]
\cP'(z_{13}) + \cS'_\a(z_{13})^2 \nn \ . \eea Summing over spin
structures only the first and the last term survive. Let us denote
them by $\cU_\cN(z_i)$ and $\cV_\cN(z_i)$. The former simply reads
\be \cU_\cN(z_i) = - \omega(z_1,z_2,z_3)\omega(z_1,z_4,z_3)
\cE_{{\cN}}(u^I_{ab}) \label{Uterm} \ . \ee The latter is more
laborious. Observing that \be
 \cS'_\a(z)^2 = \de_z [\cS_\a(z)\cS'_\a(z)] - \cS_\a(z)\cS''_\a(z)
\label{SSderiv} \ , \ee which, using (\ref{SStoP}), in turn gives
\be
 \cS'_\a(z)^2 = {1\over 2} \cP''(z) - \cS_\a(z)\cS''_\a(z)
\ee it is clear that only the second term contributes after
summation over the spin structures, so that \be \cV_\cN(z_i)
=\lim_{z_0\rightarrow z_1} \de^2_{z_0} \sum_\a c_\a
\cS_\a(z_{03})\cS_\a(z_{13}) \cZ^\cN_\a \ . \ee Further using
(\ref{SSident}) in the following guise \be
\cS_\a(z_{03})\cS_\a(z_{13}) = \cS_\a(z_{01})\left[
\omega_{z_0-z_1}(z_3) + {\theta_\a'(z_{01})\over
\theta_\a(z_{01})}\right] \ ,\ee where \be \omega_{x-y}(z) = \de_z
\log{\theta_1(z-x)\over \theta_1(z-y)} \ee is a differential of
the third kind with two simple poles with opposite residues ($\pm
1$) at $z=x$ and $z=y$, yields \be \cV_\cN(z_i) = -
\lim_{z_0\rightarrow z_1} \de^2_{z_0} \sum_\a c_\a
\left[{\theta_\a(z_{01})\over
\theta_1(z_{01})}\omega_{z_0-z_1}(z_3) + {\theta_\a(z_{01})\over
\theta_1(z_{01})}\right]   \cZ^\cN_\a \ . \ee

Both terms can be computed by means of the Riemann identity for
even spin structures \bea &&\sum_\a c_\a \theta_\a(z_1)
\theta_\a(z_2)\theta_\a(z_3)\theta_\a(z_4) = \nn \\
&&\theta_1(z'_1) \theta_1(z'_2)\theta_1(z'_3)\theta_1(z'_4) -
\theta_1(z''_1) \theta_1(z''_2)\theta_1(z''_3)\theta_1(z''_4) \ ,
\eea where $z_i'$ and $z_i''$ are related to $z_i$ through \bea
&&z_1' = {1\over 2} (z_1 + z_2 +z_3 +z_4 ) \qquad z_2' = {1\over
2}
(z_1 + z_2 -z_3 -z_4 ) \nn \\
&&z_3' = {1\over 2} (z_1 - z_2 +z_3 -z_4 ) \qquad z_4' = {1\over
2} (z_1 - z_2 -z_3 + z_4 ) \eea and \bea &&z_1'' = {1\over 2}
(-z_1 + z_2 +z_3 +z_4 ) \qquad z_2'' = {1\over 2}
(z_1 - z_2 +z_3 +z_4 ) \nn \\
&&z_3'' = {1\over 2} (z_1 + z_2 -z_3 +z_4 ) \qquad z_4'' = {1\over
2} (z_1 + z_2 +z_3 - z_4 ) \ . \eea

For ${\cN} =1$ sectors \bea &&
\cV_{\cN =1}(z_i) = - 2\pi \cX^{\cN =1}_{ab} \times \\
 && \lim_{z_0\rightarrow z_1} \de^2_{z_0} \left[ {
(\omega_{z_0-z_1}(z_3) + \de_{z_0})\{\theta_1(z_{01}/2)
[\cH(z_{01}/2) - \cH(-z_{01}/2)]\} \over \theta_1(z_{01}) \cH(0) }
\right] \nn \ , \eea where $\cH(z)$ is defined in (\ref{hfunct}),
that eventually yields \bea \cV_{\cN =1}(z_i) &=& - {2\pi}
\cX^{\cN=1}_{ab} {\cH'(0)\over \cH(0)} \left[
\de_3{\theta'_1(z_{31})\over \theta_1(z_{31})} + {1 \over 6}
{\theta'''_1(0)\over \theta'_1(0)}
- {1 \over 6}{\cH'''(0)\over \cH'(0)} \right]  \nn \\
&=&\cE_{\cN = 1} \cP(z_{13}) + \cJ_{\cN = 1} \ ,  \eea where \be
\cJ_{\cN = 1} = {2\pi} \cX^{\cN=1}_{ab} {\cH'(0)\over \cH(0)}
\left[3 \eta_1 + {1 \over 6}{\cH'''(0)\over \cH'(0)} \right]
\label{Jfunct1} \ . \ee Including the bosonic contraction
producing the momentum factor (\ref{momfact}) and adding the term
(\ref{Uterm}), one  eventually finds \bea && \la
(\psi\psi)^4\ra_{conn}^{\cN =1} = {1\over 2} (f_1 f_2 f_3 f_4)
\Pi(p_i,z_i)\{ 2\cJ_{\cN =1}+
\\ &&  + \cE_{\cN =1} [\cP(z_{13}) - \omega(z_1,z_2,z_3)
\omega(z_1,z_4,z_3) + \cP(z_{24}) -
\omega(z_2,z_1,z_4)\omega(z_2,z_3,z_4)]\}\nn \ , \eea where
$\cE_{\cN =1}$ has been defined above (\ref{Eterm1}).  Symmetry
under the exchange of $z_1,z_3$ with $z_2,z_4$ has been made
explicit though unnecessary, since we expect it to be \be
\cP(z_{31}) - \omega(z_1,z_2,z_3)\omega(z_1,z_4,z_3) = \cP(z_{24})
- \omega(z_2,z_1,z_4)\omega(z_2,z_3,z_4) \label{exchident} \ee
from consideration of periodicity and singularity.

For ${\cN} =2$ sectors, ones finds \be \cV_{\cN =2}(z_i) = -
4\pi^2 \cX_{ab}^{\cN = 2} \ \lim_{z_0\rightarrow z_1} \de^2_{z_0}
\sum_\a c_\a{\theta_\a(z_{03})\over
\theta_1(z_{03})}{\theta_\a(z_{13})\over
\theta_1(z_{13})}{\theta_\a(u_{ab})\over
\theta_1(u_{ab})}{\theta_\a(-u_{ab})\over \theta_1(-u_{ab})} \ .
\ee The limit yields \bea&&
\de^2_{z_0}\left[\theta_1\left({z_{03}+z_{13}\over 2} \right)
\theta_1\left({z_{03}+z_{13}\over 2} \right)
\theta_1\left({z_{03}-z_{13}\over 2} -u_{ab}  \right)
\theta_1\left({z_{03}-z_{13}\over 2} + u_{ab} \right) + \right. \nn \\
&& \left. \theta_1\left({z_{03}-z_{13}\over 2} \right)
\theta_1\left({z_{03}-z_{13}\over 2} \right)
\theta_1\left({z_{03}+z_{13}\over 2} -u_{ab}  \right)
\theta_1\left({z_{03}+z_{13}\over 2} + u_{ab}
\right)\right]_{z_0=z_1}
\nn \\
&& =  [\cP(z_{31}) - \cP (u_{ab})]
\theta_1(z_{13})^2\theta_1(-u_{ab})^2 \ . \eea So that, including
the momentum factor (\ref{momfact}), one eventually obtains \bea
&&\la(\psi\psi)^4\ra_{conn}^{\cN =2} = {1\over 2} (f_1 f_2 f_3
f_4) \Pi(p_i,z_i)\{2\cJ_{\cN =2} -
\\ &&
+ \cE_{\cN =2} [\cP(z_{13}) -
\omega(z_1,z_2,z_3)\omega(z_1,z_4,z_3) + \cP(z_{24}) -
\omega(z_2,z_1,z_4)\omega(z_2,z_3,z_4)]\} \nn \ , \eea where
 \be \cJ_{\cN =2} = - \cE_{\cN =2}\cP(u_{ab})
= - 4\pi^2 \cX^{\cN =2}_{ab} \cP(u_{ab}) \ . \label{Jfunct2}\ee

The other kind of disconnected contractions lead to three
inequivalent possibilities that yield terms of the form \be \la
(\ps\ps)^4\ra_{\a}^{disc} = {2^2\over 2^4} (f_1 f_2)(f_3 f_4)
\cS^2_\a(z_{12})\cS^2_\a(z_{34})\cZ_\a^\cN \ . \ee Dropping the
kinematical factor, summation over spin structures yields \be
[\cP(z_{12}) + \cP(z_{34})] \cE_\cN + \cJ_\cN \ , \ee where \be
\cJ_\cN = \sum_\a c_\a e^2_{\a-1} \cZ_\a^\cN \label{newJ} \ee
turns out to hold for the $\cJ_\cN$ previously defined in
(\ref{Jfunct1}) and (\ref{Jfunct2}). Indeed, (\ref{newJ})  can be
simplified using (\ref{SSident}) and \cite{Eguchi:1986sb} \be
e_{\a-1}^2 = i\pi\de_\tau e_{\a-1} + 2\eta_1 e_{\a-1} + {1\over 6}
g_2 \ee to \be \cJ_\cN = 2\et_1 \cE_\cN -{1\over 4} \sum_\a c_\a
\left[{\th^{(4)}_\a(0)\over \th_\a(0)} - \left({\th''_\a(0)\over
\th_\a(0)}\right)^2 \right]\cZ_\a^\cN \ . \ee

In ${\cN}=1$ sectors  one finds \be \sum_\a c_\a
{\th^{(4)}_\a(0)\over \et^3} \prod_I {\theta_\a(u^I)\over
\theta_1(u^I)} = {1\over 2} \cE_{\cN =1}\left[{ \cH'''\over \cH'}
- 6\et_1 \right] \label{theta4} \ee and \be \sum_\a c_\a
{\th''_\a(0)^2\over \et^3 \th_\a(0)} \prod_I {\theta_\a(u^I)\over
\theta_1(u^I)} = {1\over 6} \sum_\a c_\a {\de^4_z
[\th_\a(z)^2]_{z=0} - 2 \th_\a(0) \th^{(4)}_\a(0) \over \et^3
\th_\a(0)} \prod_I {\th_\a(u^I)\over \th_1(u^I)} \ . \ee The
second term is given in (\ref{theta4}), while the first can be
computed by means of the identity \be {\th_\a(z)^2 \over \th_\a(0)
} = { \cS_\a(z)^2 \th_1(z)^2 \th_\a(0) \over  \th_1'(0)^2 } = {
[\cP(z) - e_{\a-1}] \th_1(z)^2 \th_\a(0) \over \th_1'(0)^2 } \ ,
\ee that after differentiation and summation over the spin
structures yields \be {1\over 6} 2\pi \cX^{\cN = 1}_{ab} \sum_\a
c_\a {\de^4_z [\th_\a(z)^2]_{z=0} \over \et^3 \th_\a(0)} \prod_I
{\th_\a(u^I_{ab})\over \th_1(u^I_{ab})} = - 8 \et_1 \cE_{\cN =1}
\ee so that eventually  one indeed finds \be \cJ_{\cN =1} =
\cE_{\cN =1}\left[ {1 \over 6}{\cH'''(0)\over \cH'(0)} +
3\eta_1\right] \ . \label{Jterm1} \ee as above (\ref{Jfunct1}).
Including the momentum factor (\ref{momfact}) yields \be \la
(\psi\psi)^4\ra_{disc}^{\cN =1} = {1\over 4} (f_1 f_2)(f_3 f_4) \{
\cE_{\cN =1} [\cP(z_{12})+ \cP(z_{34})] + \cJ_{\cN =1} \}
\Pi(p_i,z_i) \ . \ee

In ${\cN}=2$ sectors  one has \bea &&\cJ_{\cN =2} = - {\cX^{\cN =
2} \over \et^6} \sum_\a c_\a [\th''_\a(0)
 - 2\et_1\theta_\a(0)]^2 {\theta_\a(u_{ab})^2\over \theta_1(u_{ab})^2} \nn \\
&& = 4\et_1 \cE_{\cN =2} - {\cX^{\cN = 2}_{ab}\over \et^6} \sum_\a
c_\a \th''_\a(0)^2 {\theta_\a(u_{ab})^2\over \theta_1(u_{ab})^2}
\label{thetaN2} \ . \eea The last sum leads to
 \bea&& - \de^2_{z}\de^2_w\left[\theta_1\left({z+w\over 2}
\right) \theta_1\left({z+w\over 2} \right) \theta_1\left({z-w\over
2}
-u_{ab} \right) \theta_1\left({z-w\over 2} + u_{ab} \right) + \right. \nn \\
&& \left. \theta_1\left({z-w\over 2} \right)
\theta_1\left({z-w\over 2} \right) \theta_1\left({z+w\over 2}
-u_{ab}
\right) \theta_1\left({z+w\over 2} + u_{ab} \right)\right]_{z=w=0} \nn \\
&& = - \th_1'(0)^2[\de_u^2\log\th_1(u_{ab}) - 2\et_1] =
\th_1'(0)^2 \cP(u_{ab}) \ . \eea

 Including the momentum factor finally yields \be
\la (\psi\psi)^4\ra_{disc}^{\cN =2} = {1\over 4} (f_1 f_2)(f_3
f_4) \{ \cE_{\cN =2} [\cP(z_{12})+ \cP(z_{34})] - \cJ_{\cN =2} \}
\Pi(p_i,z_i) \ , \ee where \be \cJ_{\cN =2} = - \cE_{\cN
=2}\cP(u_{ab}) \ee as above (\ref{Jfunct2}) and $\cE_{\cN =2}$ is
defined in (\ref{Eterm2}).

\section{Summary of the results for CP even amplitudes}
\label{summa}

Let us summarize our results in the NSR formalism according to the
number of supersymmetries preserved for the CP even amplitudes
receiving contribution from the sum over even spin structures.

\subsection{No fermion bilinears }

\be \la a_1\cdot\de X e^{ip_1\cdot X}(z_1) a_2\cdot\de X
e^{ip_2\cdot X}(z_2)a_3\cdot\de X e^{ip_3\cdot X}(z_3) a_4\cdot\de
X e^{ip_4\cdot X}(z_4)\ra_{even} = 0 \ee
 in any supersymmetric sector after summing over the even spin
structures.

\subsection{One fermion bilinear }

\be \la f^1_{\mu\nu}\psi^\m\psi^\n e^{ip_1\cdot X}(z_1)
a_2\cdot\de X e^{ip_2\cdot X}(z_2)a_3\cdot\de X e^{ip_3\cdot
X}(z_3) a_4\cdot\de X e^{ip_4\cdot X}(z_4) \ra_{even} = 0 \ee
 in any supersymmetric sector after summing over the even spin
structures.

\subsection{Two fermion bilinears }

\bea &&\la{i\over 2}f^1_{\mu_1\nu_1}\psi^{\m_1}\psi^{\n_1}
e^{ip_1\cdot X}(z_1){i\over
2}f^2_{\mu_2\nu_2}\psi^{\m_2}\psi^{\n_2}e^{ip_2\cdot X}(z_2)
a_3\cdot\de X e^{ip_3\cdot X}(z_3)
a_4\cdot\de X e^{ip_4\cdot X}(z_4)\ra_{even} \nn  \\
&& = -{1\over 2} (f_1 f_2) {\cal E}_{\cN} \Pi(z_i; p_i)
\left[a_3\cdot a_4 \de_3 \de_4 \cG_{34} - \sum_{i\neq 3} a_3\cdot
p_i \de_3 \cG_{3i}\sum_{j\neq 4}a_4\cdot p_j \de_4 \cG_{4j}\right]
\nn \eea plus permutations (6 in all), where \be \Pi(p_i) =
\prod_{i<j} \exp(-p_i\cdot p_j \cG_{ij}) \ee and, depending on the
number of supersymmetries $\cN$,  \be {\cal E}_{{\cN} = 4} = 0
\quad , \qquad {\cal E}_{{\cN} = 2} = (2\pi)^2 {\cal X}^{\cN =
2}_{ab}  \quad , \qquad {\cal E}_{{\cN} = 1} = 2\pi {\cal X}^{\cN
= 1}_{ab} {\cH'(0)\over \cH(0) } \ , \ee with $\cH(z) = \prod_I
\theta_1(z + u^I_{ab})$ and,  up to $\delta(\sum_i p_i)$,\be {\cal
X}^{\cN = 4}_{ab} = {(2\pi)^4 \Lambda_{ab}\over 4 n (\ap t)^2}  \
, \qquad  {\cal X}^{\cN = 2}_{ab} = {(2\pi)^4 {\cal
I}^{\perp}_{ab} {\Lambda}^{\parallel}_{ab} \over 4 n (\ap t)^2}  \
, \qquad  {\cal X}^{\cN = 1}_{ab} = {(2\pi)^4 {\cal I}_{ab} \over
4 n (\ap t)^2}  \ . \ee

\subsection{Three fermion bilinears }

\bea &&\la{i\over 2}f^1_{\mu_1\nu_1}\psi^{\m_1}\psi^{\n_1}
e^{ip_1\cdot X}(z_1) .. {i\over
2}f^3_{\mu_3\nu_3}\psi^{\m_3}\psi^{\n_3}e^{ip_3\cdot X}(z_3)
a_4\cdot\de X e^{ip_4\cdot X}(z_4)\ra_{even} \nn  \\ && = (f_1 f_2
f_3) {\cal E}_{\cN} \omega(z_1,z_2,z_3) \sum_{j\neq 4}a_4\cdot p_j
\de_4 \cG_{4j} \Pi(z_i;p_i)\nn \eea plus permutations (4 in all)
where \be \omega(z_1,z_2,z_3) = \de_1\log\theta_1(z_{12}) +
\de_2\log\theta_1(z_{23}) + \de_3\log\theta_1(z_{31}) = \de_1
\cG_{12} + \de_2  \cG_{23} + \de_3  \cG_{31} \ . \ee

\subsection{Four fermion bilinears, connected}
 \bea
&&\la {i\over 2}f^1_{\mu_1\nu_1}\psi^{\m_1}\psi^{\n_1}
e^{ip_1\cdot X}(z_1) ... {i\over
2}f^4_{\mu_4\nu_4}\psi^{\m_4}\psi^{\n_4}e^{ip_4\cdot
X}(z_4)\ra_{even}^{conn} = {1\over 2} (f_1 f_2f_3 f_4) \times \nn
\\&& \Pi(z_i;p_i) [{\cal E}_{\cN} [\cP(z_{13})
-\omega(z_1,z_2,z_3)\omega(z_1,z_4,z_3)] + {\cal J}_{\cN} + (1,3
\leftrightarrow 2,4)] \eea plus permutations (3 in all) where \be
{\cal J}_{{\cN} = 4} = (2\pi)^4 {\Lambda}^{\parallel}_{ab} \ ,
\qquad {\cal J}_{{\cN} = 2} =  - {\cal E}_{{\cN} = 2}\cP(u_{ab})\
, \qquad {\cal J}_{{\cN} = 1} = {\cal E}_{{\cN} = 1} \left[
3\eta_1 + {1\over 6}{\cH'''(0)\over \cH'(0)}\right] \ , \ee with
$\eta_1 = -2\pi i \de_\tau\log\eta$.

\subsection{Four fermion bilinears, disconnected}
 \bea &&\la {i\over 2}f^1_{\mu_1\nu_1}\psi^{\m_1}\psi^{\n_1}
e^{ip_1\cdot X}(z_1) ... {i\over
2}f^4_{\mu_4\nu_4}\psi^{\m_4}\psi^{\n_4}e^{ip_4\cdot X}(z_4)
\ra_{even}^{disc} = \nn \\
&& {1\over 4}(f_1 f_2)(f_3 f_4) \Pi(z_i;p_i) \{ {\cal E}_{\cN}
[{\cal P}(z_{12}) + {\cal P}(z_{34})]  - {\cal J}_{\cN} \} \eea
plus permutations (3 in all).

\section{CP odd amplitudes in $\cN= 1$ sectors}
\label{CPodd}

 In the odd spin structure, the presence of a
supermodulus requires the insertion of
$\delta(\beta)=e^{+\varphi}$ in order to absorb the zero mode of
the anti-superghost $\beta= e^{-\varphi}\partial\xi$. The presence
of a conformal Killing spinor  requires the insertion of
$\delta(\gamma)=e^{-\varphi}$ in order to absorb the zero mode of
the superghost $\gamma= \eta e^{+\varphi}$. This allows one to fix
the position in superspace of one of the vertices that would than
be of the form $V=a\cdot\psi \exp(ipX)$. The two  combined
operations are equivalent to inserting a picture changing operator
$\Gamma = e^{\varphi} G + ...$, where $G$ is the worldsheet
supercurrent, at an arbitrary point $z_0$ and using the (-1)
picture for one of the vertices. Independence from $z_0$ allows
one to let $z_0$ coincide with the position of the vertex in the
(-1) picture and replace it with the expression in the (0) picture
after using $\la e^{\varphi(z)} e^{-\varphi(w)}\ra_{odd} =1 $.
Moreover one has to absorb the four zero-modes of the spacetime
fermions present in ${\cN} =1$ sectors. In $\cN =2$ and $\cN =4$
sectors, CP odd amplitudes with only vector bosons vanish. There
is no way to absorb the two (for $\cN =2$) or six (for $\cN =4$)
additional zero-modes of the internal fermions present in these
sectors. More complicated amplitudes with matter scalars and
fermions can accomplish the task.

Let us thus concentrate on ${\cN} =1$ sectors and start from the
simplest non vanishing contribution.

\subsection{Two fermion bilinears (6 terms)}

Thanks to the exact cancellation between bosonic and fermionic
non-zero modes on the worldsheet the final result for terms of the
form $\la \de X \de X :\ps\ps: :\ps\ps: \ra_{odd}$ is very simple
and compact \bea &&\la \de X(z_1) \de X(z_2)
:\ps\ps:(z_3):\ps\ps:(z_4)\ra_{odd} = {2\over
2^2}\left(\sqrt{2\over \tau^\cA_2}\right)^4 (f_3\cdot\tilde{f}_4)
\cX^{\cN =1}_{ab} \times
 \nn \\ && [ a_1\cdot a_2 \de_1\de_2 \cG(z_{12}) -
\sum_{i\neq 1} a_1\cdot p_i \de_1 \cG(z_{1i}) \sum_{j\neq 2}
a_2\cdot p_j\de_2 \cG(z_{2j})] \Pi(z_i;p_i) \eea where
$\Pi(z_i;p_i)$ denotes the momentum factor (\ref{momfact}) and the
overall coefficient takes into account symmetry factors and the
correct normalization of the fermionic zero-modes. In addition
there are five more permutations.

\subsection{Three fermion bilinears (12
terms)}

The next simplest term is \be \la :\ps\ps:(z_1)
:\ps\ps:(z_2):\ps\ps:(z_3)\de X(z_4)\ra_{odd} \ . \ee  The four
zero-modes can be absorbed in three distinct ways. For instance,
absorbing two of them at $z_3$, one at $z_1$ and one at $z_2$ and
contracting the remaining two fermions at $z_1$ and $z_2$ yield
\be \ \la \de X (\ps\ps)^3 \ra_{odd} = {2^3\over
2^3}\left(\sqrt{2\over \tau^\cA_2}\right)^4 (f_1 \cdot f_2 \cdot
\tilde{f}_3) \cX^{\cN =1}_{ab} \cS(z_{12}) \Pi(p_i; z_i)
\sum_{i\neq 4} i a_4\cdot p_i \de_4 \cG(z_{i4}) \ee plus two more
permutations. In the odd spin structure the fermionic propagator
can be taken to be \cite{Stieberger:2002fh,Stieberger:2002wk} \be
\cS(z-w) = - \de_z\cG(z-w) = \de_z \log\theta_1(z-w) + 2\pi i
{Im(z-w) \over Im\tau} \  . \ee

\subsection{Four fermion bilinears (21
terms, 3 structures)}

Finally the most laborious term is \be \la :\ps\ps:(z_1)
:\ps\ps:(z_2):\ps\ps:(z_3):\ps\ps:(z_4)\ra_{odd} \ . \ee  In this
case there are three possible ways of absorbing zero-modes.

Absorbing two zero-modes at one point (say $z_1$) and two at
another point (say $z_2$), for a total of 6 permutations,
contributes expressions of the form \be \la (\ps\ps)^4
\ra^{(2_02_000)}_{odd} = - {2^2\over 2^4} \left(\sqrt{2\over
\tau^\cA_2}\right)^4 (f_1\tilde{f}_2) (f_3\cdot f_4)\cX^{\cN
=1}_{ab} \cS^2(z_{34}) \Pi(p_i; z_i) \ee plus permutations.

Next, one can absorb two zero-modes at one point (say $z_1$), one
at another point (say $z_2$), and one at a third point (say $z_3$)
for a total of 12 permutations contributing expressions of the
form \be \la (\ps\ps)^4 \ra^{(2_01_01_00)}_{odd} = {2^4\over 2^4}
\left(\sqrt{2\over \tau^\cA_2}\right)^4 (f_2\cdot\tilde{f}_1 \cdot
f_3\cdot f_4) \cX^{\cN =1}_{ab}\cS(z_{24})\cS(z_{34}) \Pi(p_i;
z_i)  \ee plus permutations.

Finally one can absorb one zero mode at each point which yields
\be \la (\ps\ps)^4 \ra^{(1_01_01_01_0)}_{odd} = {2^4\over 2^4}
\left(\sqrt{2\over \tau^\cA_2}\right)^4
\e_{{\m_1}{\m_2}{\m_3}{\m_4}} f_1^{{\m_1} {\n}}{f}_2^{\m_1}{}_\n
f_3^{{\m_3} {\r}}{f}_4^{\m_4}{}_\r \cX^{\cN =1}_{ab}
\cS(z_{12})\cS(z_{34}) \Pi(p_i; z_i) \ , \ee as well as two more
permutations arising from different Wick contractions of the
fermionic non-zero modes.

\section{Forward scattering}
\label{forwscatt}

 The recipe for computing string amplitudes
requires integrating over the insertion points and then over the
modular parameter(s) of the relevant Riemann surface. The task is
prohibitively complicated, if not impossible, in general. Yet for
some very special amplitudes or kinematic regimes the situation
drastically simplifies. This is the case for non-planar forward
scattering that, as we will see, allows to extract interesting
predictions for near future colliders.

For forward scattering $p_1=-p_4$, $p_2 = -p_3$. As a result there
is only one non-zero kinematical invariant \be p_1\cdot p_2 =
p_3\cdot p_4 = - p_1\cdot p_3 = p_2\cdot p_4 = -s/2 = u/2 \quad
p_1\cdot p_4 = p_2\cdot p_3 = 0 = t/2 \ee so that \be \Pi(p_i,
z_i) \rightarrow \Pi(s, z_i) = \exp(s/2(\cG_{12}- \cG_{13}+
\cG_{34}- \cG_{24}) \ . \ee

Moreover, since $a_1 = a_4$ and $a_2 = a_3$, one easily finds \be
(f_1 f_1) = (f_4 f_4) = - (f_1 f_4) = 0 \quad , \quad (f_2 f_2) =
(f_3 f_3) = - (f_2 f_3) = 0 \ee and \bea &&(f_1 \tilde{f}_1) =
(f_4
\tilde{f}_4) = -  (f_1 \tilde{f}_4) = (f_4 \tilde{f}_1) = 0 \\
&&(f_2 \tilde{f}_2) = (f_3 \tilde{f}_3) = - (f_2 \tilde{f}_3) =
(f_3 \tilde{f}_2) = 0 \ , \eea in addition all cubic contractions
vanish \be (f_i f_j f_k) = 0 \quad , \quad (f_i f_j \tilde{f}_k) =
0 \ee since at least two of the $f$'s are equal (opposite). As a
result contractions involving three fermion bilinears give
vanishing contribution to forward scattering both to CP even and
CP odd processes.

Moreover for non-planar amplitudes the two stacks of branes should
be of the same kind $a=b$ so that $T_1 = T_4^\dagger$ and $T_2 =
T_3^\dagger$ and $tr_a(T_1T_2) = tr_b(T_3T_4)$.

Notice that integration over the four points is unrestricted in
the non-planar case, since the Chan-Paton factor $\tr(T_1 T_2)
\tr(T_3 T_4)$ is invariant under re-ordering of $z_1,z_2$ and of
$z_3,z_4$. So even if {\it a priori} $0<z_1<z_2<1$ for a given
Chan-Paton factor $\tr(T_1 T_2) \tr(T_3 T_4)$, the other ordering
$0<z_2<z_1<1$ has the same Chan-Paton factor since $\tr(T_1 T_2)=
\tr(T_2 T_1)$. This extends immediately to twisted sectors whereby
$\tr(T_1 T_2 W^k) \tr(T_3 T_4 W^k)$ is also invariant under
reordering since $[W, T_i]=0$. Integrating by parts is thus
possible and further simplifies the non-planar forward scattering
amplitudes.

\subsection{CP even amplitudes}

For forward scattering, dropping total derivatives and Chan-Paton
factors but including all relevant permutations, CP even
amplitudes with two fermion bilinears read
 \bea &&\la{i\over 2}f^1_{\mu_1\nu_1}\psi^{\m_1}\psi^{\n_1}
e^{ip_1\cdot X(z_1)}{i\over
2}f^2_{\mu_2\nu_2}\psi^{\m_2}\psi^{\n_2}e^{ip_2\cdot X(z_2)}
a_2\cdot\de_3 X e^{-ip_2\cdot X(z_3)}
a_1\cdot\de_4 X e^{-ip_1\cdot X(z_4)}\ra^{FS}_{even} \nn  \\
&& + {\rm perms} = -{(f_1 f_2)^2 \over 2\ap s} {\cal E}_{\cN}
[\de_1 \de_2 \cG_{12} + \de_3 \de_4 \cG_{34} + \de_1 \de_3
\cG_{13} + \de_2 \de_4 \cG_{24}] \Pi(z_i; s)\nn \ , \eea where,
for shortness, $\cG_{ij} = \cG(z_{ij})$ and \be (f_1 f_2) = 2
[(a_1\cdot p_2)(a_2\cdot p_1)- (a_1\cdot a_2)(p_1\cdot p_2)] \ .
\ee

CP even amplitudes arising from connected (box-type) contractions
of four fermion bilinears can be conveniently simplified using the
identity \cite{Stieberger:2002fh,Stieberger:2002wk} \be
\cB_\a(z_1,z_2,z_3,z_4) +
\cB_\a(z_1,z_3,z_2,z_4)+\cB_\a(z_1,z_3,z_4,z_2) = {1\over 2}
\de_z^4 \log\theta_\a(z)|_{z=0} \ , \ee where \be
\cB_\a(z_1,z_2,z_3,z_4) =
\cS_\a(z_{12})\cS_\a(z_{23})\cS_\a(z_{34})\cS_\a(z_{14})  \ee and
observing that for forward scattering \be t_1 = (f_1 f_2 f_3 f_4)
= (f_1 f_2 f_2 f_1) = (f_1 f_3 f_2 f_4) = a_1^2 a_2^2 (p_1\cdot
p_2)^2  \ee and \be t_2 = (f_1 f_3 f_4 f_2) = (f_1 f_2 f_1 f_2) =
{1\over 2}(f_1 f_2)^2 = 2 [(a_1\cdot p_2)(a_2\cdot p_1)- (a_1\cdot
a_2)(p_1\cdot p_2)]^2 \ . \ee One then has
 \bea
&&\{ t_1[\cB_\a(z_1,z_2,z_3,z_4) + \cB_\a(z_1,z_3,z_2,z_4)] + t_2
\cB_a(z_1,z_3,z_4,z_2)\}\cZ_\a  \nn \\ && = {1\over 2} t_1 \cZ_\a
\de_z^4 \log\theta_\a(z)|_{z=0} + (t_2 - t_1)
\cB_a(z_1,z_3,z_4,z_2) \cZ_\a \ . \eea Summing over the spin
structures eventually yields \bea &&\la{i\over
2}f^1_{\mu_1\nu_1}\psi^{\m_1}\psi^{\n_1} e^{ip_1\cdot
X(z_1)}{i\over
2}f^2_{\mu_2\nu_2}\psi^{\m_2}\psi^{\n_2}e^{ip_2\cdot X(z_2)} \nn
\\
&& {i\over 2}f^2_{\mu_3\nu_3}\psi^{\m_3}\psi^{\n_3}e^{-ip_2\cdot
X(z_3)} {i\over
2}f^1_{\mu_4\nu_4}\psi^{\m_4}\psi^{\n_4}e^{-ip_4\cdot
X(z_4)}\ra^{FS,conn}_{even} =  \\ &&\Pi(z_i; s) \{
(t_2-t_1)\cE_\cN [\cP(z_{14}) -
\omega(z_1,z_4,z_2)\omega(z_1,z_4,z_3)] + (t_2 + 2t_1) \cJ_\cN\}
\nn \ . \eea

Recall that symmetry under $(14)\leftrightarrow (23)$ exchange is
expected and would follow if (\ref{exchident}) hold.

For forward scattering the disconnected CP even contractions of
four fermion bilinears yield \bea &&\la{i\over
2}f^1_{\mu_1\nu_1}\psi^{\m_1}\psi^{\n_1} e^{ip_1\cdot
X(z_1)}{i\over
2}f^2_{\mu_2\nu_2}\psi^{\m_2}\psi^{\n_2}e^{ip_2\cdot X(z_2)} \nn
\\
&& {i\over 2}f^2_{\mu_3\nu_3}\psi^{\m_3}\psi^{\n_3}e^{-ip_2\cdot
X(z_3)} {i\over
2}f^1_{\mu_4\nu_4}\psi^{\m_4}\psi^{\n_4}e^{-ip_4\cdot
X(z_4)}\ra^{FS,disc}_{even} =  \\ && {1\over 2} \Pi(z_i; s) t_2 \{
\cE_\cN [\cP(z_{12}) + \cP(z_{34})+ \cP(z_{13}) + \cP(z_{24})] - 2
\cJ_\cN\} \nn \ . \eea

\subsection{CP odd amplitudes}

For forward scattering CP odd contractions of two fermion
bilinears, after integrating by parts and including all the four
non-vanishing permutations, one finds (recall $\ap = 1/2$) \bea &&
\ \la \de X \ \de X :\psi\psi: :\psi\psi: \ra^{FS}_{odd} = - {(f_1
f_2)(f_1 \tilde{f}_2)\over 4\ap s }\left(\sqrt{2\over
\tau^\cA_2}\right)^4 \cX_{ab}^{\cN = 1}
\times \\
&& \Pi(z_i; s) [\de_1\de_2 \cG(z_{12}) + \de_3\de_4 \cG(z_{34}) +
\de_1\de_3 \cG(z_{13}) + \de_2\de_4 \cG(z_{24})] \nn \ . \eea

For forward scattering CP odd contractions of four fermion
bilinears can be simplified by means of the identities \be
\e_{{\m_1}{\m_2}{\m_3}{\m_4}} f_1^{{\m_1}
{\n}}\et_{\n\l}{f}_2^{{\m_1} {\l}} f_1^{{\m_3}
{\r}}\et_{\r\s}{f}_2^{{\m_4} {\s}} = - {1\over 2} (f_1
\tilde{f}_2) (f_1 f_2) \ee and \be (f_1 \tilde{f}_2  f_1  f_2) =
{1\over 2} (f_1 \tilde{f}_2) (f_1 f_2) \ . \ee

Thus eventually one gets \be \la :\psi\psi: \ :\psi\psi: \
:\psi\psi: \ :\psi\psi: \ra^{FS}_{odd} = {1\over 4} (f_1
\tilde{f}_2) (f_1 f_2) \cX_{ab}^{\cN = 1} (\cS_{12} - \cS_{34} -
\cS_{13} + \cS_{24})^2  \ . \ee

\section{Imaginary part and total cross section}
\label{totcross}

 According to the optical theorem, the total cross section for the
production of closed string states in the bulk, obtains from the
imaginary part of the non-planar forward scattering amplitude \be
\s_{tot}(s) = {1\over s} Im\cA_{FS}(s) \ . \ee

It turns out to prove convenient to transform the non-planar
amplitude to the transverse channel that exposes the `tree-level'
unoriented closed string exchange. It is remarkable but not
unexpected that our final compact expressions for the amplitudes
transform covariantly, thus providing a check of their validity,
if needed. Indeed, performing an S-modular transformation from
$\tau = it/2$ to $\tilde\tau= i\ell)$ one finds \be \cF_\cN(\tau
=-1/\tilde\tau) = -i\tilde\tau \cF_\cN(\tilde\tau) \ee for all the
correlators we have computed in any sector of the theory. The
overall power of $\tilde\tau$ then cancels against the measure of
integration $dt/t = d\ell/\ell$. Under the required S-modular
transformation, the combinations $u^I_{ab} = k v^I_{ab} +
\ve^I_{ab} \tau_\cA$ transform into $\tilde{u}^I_{ab} = k
v^I_{ab}\tilde\tau_\cA - \ve^I_{ab}$. What was a projection in the
direct channel becomes a mass-shift in the transverse channel and
{\it vice versa}. The boundary insertion points $z_1$ and $z_2$
gets re-located onto a unit segment along the real axis, while
$z_3$ and $z_4$ gets re-located onto a unit segment parallel to
the real axis and displaced from it by an amount
$\tilde\tau_2/2=\ell/2$.

Thanks to the symmetry of the Chan-Paton factors, the integration
is unrestricted and total derivatives can be dropped since there
is no boundary contribution. Indeed, terms of the form \be \de_1
[\cG(z_{12}) - \cG(z_{12})] \de_4 [\cG(z_{42}) - \cG(z_{43})]
\Pi(s,z_i) = (2/s)^2 \de_1 \de_4 \Pi(s,z_i) \ee being total
derivatives integrate to zero. The relative  sign appear due to
$p_3=-p_2$. Similarly for $2,3$ since $p_4=-p_1$.

Terms of the form \be \de_1 [\cG(z_{12}) - \cG(z_{13})] \de_2
[\cG(z_{21}) -  \cG(z_{24})] \Pi(s,z_i)\ee are more involved. One
has \bea &&\de_1 [\cG(z_{12}) - \cG(z_{13})] \de_2 [\cG(z_{21}) -
\cG(z_{24})] \Pi(s,z_i) = (2/s) \de_1 [\cG(z_{12}) - \cG(z_{13})]
\de_2 \Pi(s,z_i) \nn \\ &&= (2/s) \de_2 \{ \de_1 [\cG(z_{12}) -
\cG(z_{13})]\Pi(s,z_i)\} - (2/s) \de_2\de_1 [\cG(z_{12}) -
\cG(z_{13})] \Pi(s,z_i) \ . \eea The first term is a total
derivative and integrates to zero. The second term can be
conveniently rewritten as \be \de_2 \de_1 \cG(z_{12}) \Pi(s,z_i) =
-{\ap \over 2} [\cP(z_{12}) + 2\eta_1 + {\pi \over 2\tau_2^\cA}]
\Pi(s,z_i) \ . \ee The same applies to the pairs of points
$(1,3)$, $(4,3)$ and $(2,4)$.

Dropping all the tildes for simplicity, the final form of the
worldsheet integrals one needs to compute for the CP even case is
\bea && \cA^{FS}_\cN(s) = \int d\ell \cX_\cN \int \prod_i dz_i
e^{{s\over 2} (\cG_{12} - \cG_{13}+ \cG_{34}- \cG_{24})} \{ 2 t_1
\cJ_\cN + {t_2\over \ap s} \cE_\cN \left( 4\eta_1 + {\pi\over
\ell} \right) + \nn
\\
&& + (t_2 - t_1) \cE_\cN [\cP_{14} - (\de_1\cG_{14})^2] +
{t_2\over 2} {\ap s + 1 \over \ap s}  \cE_\cN
[\cP_{12}+\cP_{34}+\cP_{13}+ \cP_{24}] \} \ . \eea

For $\cN = 4 $ sectors, $\cE_{\cN =4}=0$ and  only the first term
contributes and yields \be \cA^{FS}_{\cN = 4} = K_{\cN = 4}(s;
a_1, a_2) \int_0^\infty d\ell {\tilde{\Lambda}_{6-\hat{d}}(\ell)
\over \ell^{\hat{d}/2}} \prod_i \int_0^1 dx_i\Pi_X(\ell; z_i= x_i
+ i \delta_i) \ , \ee where \be K_{\cN = 4}(s; a_1, a_2) = {2
a_1^2 a_2^2 (p_1\cdot p_2)^2 (2\pi)^4 \hat{V}_{6-\hat{d}} \over
2_{GSO} 2_{\Omega} 2^2_{P_4} 2^3_{\Lambda_{6-\hat{d}}} N_{orb}}
\tr(T_1 T_2)^2  \ , \ee $0\le \hat{d}\le 6$ is the number of
`large' internal dimensions in the $D3$-brane description, \ie
`small' in the T-dual $D9$-brane description, and $\delta_1 =
\delta_2 = 0$ while $\delta_3 = \delta_4 = \ell/2$. Dependence on
the insertion points is only through $\Pi_X(\ell; z_i)$.

Exploiting the series expansions in $q = e^{-2\pi\ell}$ collected
in an appendix, one finds
\bea &&[4q^{1/4} \sin(\pi x_{12})
\sin(\pi x_{34})]^{\ap s} \Pi_X(q)
 = 1 - 2\ap s q^{1/2}[\cos(2\pi x_{13}) + \cos(2\pi x_{24})] \nn \\
&& + 2\ap s q \{1 + \cos(2\pi x_{12}) + \cos(2\pi x_{34}) + \nn
\\
&&(\ap s -1)[\cos^2(2\pi x_{13}) + \cos^2(2\pi x_{24})] + 2\ap s
\cos(2\pi x_{13})\cos(2\pi x_{24})\} \nn \\
&&+ {4\over 3} \ap s q^{3/2} \{ [1 + \cos(2\pi x_{12}) + \cos(2\pi
x_{34}) + (\ap s -1) \cos(2\pi x_{13}) \cos(2\pi x_{24})]\times
\nn \\
&& \times 3\ap s [\cos(2\pi x_{13})+ \cos(2\pi x_{24})] + (\ap s
-1)(\ap s -2)[\cos^3(2\pi x_{13}) + \cos^3(2\pi x_{24})] \} \nn \\
&&+{1\over 3} \ap s q^2 \{ 2(\ap s -1)(\ap s -2)(\ap s
-3)[\cos^4(2\pi x_{13}) + \cos^4(2\pi x_{24})] + \nn \\
&&+8\ap s \cos(2\pi x_{13}) \cos(2\pi x_{24})
[(\ap s -1)(\ap s -2)(\cos^2(2\pi x_{13}) + \cos^2(2\pi x_{24})) + \nn \\
&&3\ap s (1 + \cos(2\pi x_{12}) + \cos(2\pi x_{34}))] + \nn \\
&&+ 12 \ap s (\ap s -1)^2 \cos^2(2\pi x_{13}) \cos^2(2\pi x_{24})+ \nn \\
&&+ 12 [\cos^2(2\pi x_{13}) + \cos^2(2\pi x_{24})](1 + \ap s (\ap
s -1) [1 + \cos(2\pi x_{12}) + \cos(2\pi x_{34})]) + \nn \\
&&+ 3 \{ 2 (\ap s + 1) [ \cos(2\pi x_{12}) + \cos(2\pi x_{34})]^2
- 4\cos(2\pi x_{12})\cos(2\pi x_{34}) + \nn \\
&&+  2(2 \ap s +1)[ \cos(2\pi x_{12}) + \cos(2\pi x_{34})] + 2 \ap
s -3 \}\nn + ... \ .
\\ \eea

Truncating to lowest order (\ie $q^0$), performing the
trigonometric integrals over the insertion points by means
of\footnote{It is easy too see that $
 \int_0^1 dx (\sin\pi x)^a (\cos\pi x)^{2n+1} =0$ or more generally
$$
 \int_0^1 dx (\sin\pi x)^a (\cos\pi x)^{b} = {1 + e^{i\pi b} \over 2\pi}
{\G\left({1 + a \over 2} \right) \G\left({1 + b \over 2} \right)
\over \G\left(1+ {a + b \over 2} \right) } $$.} \be
 \int_0^1 dx (\sin\pi x)^a (\cos\pi x)^{2n} =
{\G\left({1 + a \over 2} \right) \G\left({1\over 2} + n  \right)
\over \pi \G\left(1+n + {a \over 2} \right) }  \label{trigint} \ ,
\ee and extracting the imaginary part by means of \be Im\left(
\int_0^\infty d\ell \ell^{-\a} e^{-\b \ell} \right) = {\pi
\b^{\a-1} \over \G (\a)} \label{Impart} \ , \ee one gets
\be\sigma_0(s) = K_{\cN = 4}(s; a_1, a_2){\pi \over s
\Gamma({\hat{d}\over 2})} \left( - { \pi \ap s \over
2}\right)^{{\hat{d}\over 2} - 1} \left[ {\Gamma({1 -\ap s \over
2}) \over 2^{\ap s} \sqrt{\pi} \Gamma(1 -{\ap s \over 2})}
\right]^2 \label{formfact0} \ee for $0\le \ap s < 4$, in perfect
agreement with the results of \cite{Chialva:2005gt}, {\it mutatis
mutandis}.

Integration over $x_{13}$ (equivalently $x_{24}$) effectively
kills all half odd integer powers of $q $ in the expansion of
$\Pi_X$, given in an appendix. The next contribution in $\cN = 4$
sectors thus comes from terms of order $q^1$. Performing the
trigonometric integrals over the insertion points by means of
(\ref{trigint}) and extracting the imaginary part yields

\be\sigma_1(s) = K_{\cN = 4}(s; a_1, a_2){\pi \over s
\Gamma({\hat{d}\over 2})} \left( - { \pi (\ap s - 4) \over
2}\right)^{{\hat{d}\over 2} - 1} \left[ {\Gamma({1 -\ap s \over
2}) \over 2^{\ap s} \sqrt{\pi} \Gamma(1 -{\ap s \over 2})}
\right]^2 2 B_1(\ap s) \ee for $4\le \ap s < 8$, where the 'form
factor' is given by \be B_1(\ap s) = 2 (\ap s)^2 {(\ap s)^2 - 3\ap
s + 4 \over (\ap s - 2)^2 } = 2 (\ap s)^2 {\left(\ap s - {3\over
2}\right)^2 + {7\over 4} \over (\ap s - 2)^2 }  \ , \ee in perfect
agreement with the results of \cite{Chialva:2005gt}, {\it mutatis
mutandis}.

For $\cN =2$ sectors the last term $\cJ_{\cN =2}$ gives similar
results after replacing the overall kinematical factor with \be
K_{\cN = 2}(s; a_1, a_2) = {\pi^2 \over 3} {2 a_1^2 a_2^2
(p_1\cdot p_2)^2 (2\pi)^2 \cI^\perp_{ab} \hat{V}_{2-\hat{d}} \over
2_{GSO} 2_{\Omega} 2^2_{P_4} 2_{\Lambda^\parallel_{2-\hat{d}}}
N_{orb}} \tr(W_k T_1 T_2)^2 \ , \ee where the first factor comes
from the constant term ($q^0$) in the expansion of $-\cP(u)$, and
taking into account that the number of `large' internal dimensions
satisfies $0\le \hat{d}\le 2$ in these sectors.

For $\cN =1$ sectors the last term $\cJ_{\cN =1}$ gives poles
rather than a cut since $\hat{d}=0$ in these sectors. The massless
poles corresponds to the dependence of the gauge couplings on the
VEV's of the massless closed string scalars in the twisted
sectors. Massive poles signal the possibility of producing
(unstable) closed string 'resonances' with not necessarily integer
masses (in $1/\ap$ units).

A potential negative mass pole might arise from $\cP(z_{12})$ and
similar terms. By OPE considerations however it should be absent.
Reassuringly one finds this kind of terms with coefficient
proportional to $1 + (1/\ap s)$ so that \be \left( 1 + {1\over \ap
s} \right) \int_0^1 (\sin\pi x)^{-\ap - 2} dx = {\Gamma({1 -\ap s
\over 2}) \over 2^{\ap s} \sqrt{\pi} \Gamma(1 -{\ap s \over 2})} \
 \ee has no `tachyonic' pole. The next term in the $q$ expansion of $P(z_{12})$ is a
constant ($-\pi^2/3$) that yields integrals of the form
(\ref{formfact0}). \ie the form factor of the `zero' mass states.
The term of order $q$ is proportional to $\sin^2(2\pi x)$.
Integration then yields \be \int_0^1 (\sin\pi x)^{-\ap + 2} dx =
{\ap s - 1 \over \ap s - 2} {\Gamma({1 -\ap s \over 2}) \over
2^{\ap s} \sqrt{\pi} \Gamma(1 -{\ap s \over 2})} \ . \ee

The same situation prevails for the terms in $P(z_{34})$.

Finally one should consider the combinations $\cP(z_{14}) -
(\de_1\cG_{14})^2$,  plus the corresponding ones with $(1,4)$
replaced by $(2,3)$. Quite remarkably \be \cP(z_{14}) -
 (\de_1\cG_{14})^2 = - 4 \pi i {\de \over \de \tau}
\log\left(\theta_4(x_{14}|\tau) \over \eta(\tau)\right) \ee that
admits the following expansion \bea &&\cP(z_{14}) -
 (\de_1\cG_{14})^2 = - {\pi^2 \over 3} - 8\pi^2 \sum_{n,d_n|n} { 1
- (-)^{n/d_n} \over 2} q^{n/2} {n\over d_n} \cos(2\pi d_n x_{14})
\nn \\
&& = - {\pi^2 \over 3} - 8\pi^2 [q^{1/2} \cos(2\pi x_{14}) + q
\cos(4\pi x_{14}) + ...] \ , \eea where $d_n | n$ denotes the
divisors of $n$.  Combining with $\Pi_X$ one finds that half odd
integer powers of $q$ vanish after integration over $dx_{13}$ or,
equivalently, $dx_{24}$. However new integer powers are generated
by combinations of integer and half odd integer powers of $q$
$\Pi_X$ and $\cP(z_{14}) - (\de_1\cG_{14})^2$. For instance at
order $q$ one finds \bea &&- {2\pi^2\ap s \over 3}  \{1 +
\cos(2\pi x_{12}) + \cos(2\pi x_{34}) + \nn
\\
&&(\ap s -1)[\cos^2(2\pi x_{13}) + \cos^2(2\pi x_{24})] + 2\ap s
\cos(2\pi x_{13})\cos(2\pi x_{24})\} \nn \\
&&- 8\pi^2 \cos(4\pi x_{14}) + 16 \pi^2\ap s \cos(2\pi
x_{14})[\cos(2\pi x_{13}) + \cos(2\pi x_{24})] \ . \eea After
integration over $dx_{13}$ or, equivalently, $dx_{24}$, the last
term gives $8 \pi^2\ap s [\cos(2\pi x_{34}) + \cos(2\pi x_{12})]$,
that modifies $\sigma_1(s)$ when combined with the lowest order
term in\footnote{For simplicity we assume $k v^I_{ab} < 1$.}
 \be
\cE_{\cN=1} \approx 3\pi + 2\pi \sum_I q^{k
v^I_{ab}} + 4\pi \sum_I q^{1 - k v^I_{ab}} +
... \ . \ee

New thresholds with fractional mass appear due to the fractional
powers in the expansion of $\cE_\cN$.

Terms in $\cP(z_{13})$ and $\cP(z_{24})$ can be discussed
similarly.

\section{Comments}
\label{comm}

In their present form, our results are not directly related to
processes observable at LHC. Without some recoiling observable
(open string) states it is impossible to detect the decay into
closed strings in the bulk. Yet it should not be difficult to
include some soft observable particle along the lines of
\cite{Chialva:2005gt}. For hadronic colliders, such as LHC, a much
subtler issue is how to extract hadronic cross sections from the
`partonic' cross sections we have computed. One has to convolute
our or similar results with the partonic distributions of the
relevant hadrons, \ie the proton. To the best of our knowledge
these are not known in analytic form but significant effort
\cite{workshop} is presently devoted into this important step.

At a more formal level, our results, obtained for a specific yet
interesting class of supersymmetric models with open and
unoriented strings, display a remarkably simple structure. This is
largely due to the already observed fact that open string gauge
bosons belong to the `identity' sector of the internal conformal
field theory, describing the compactification from $D=10$. We thus
see no major obstacle in extending them to the case of genuinely
interacting internal $\cN =2$ SCFT's, such as Gepner models
\cite{Angelantonj:1996mw}. It is tantalizing to speculate that \be \cE^{(s)} = -
\sum_\a c_\a e_{\a-1} \cZ^{(s)}_\a \ee and \be \cJ^{(s)} = \sum_\a
c_\a e^2_{\a-1} \cZ^{(s)}_\a \ee should remain valid, once the
relevant partition functions $\cZ^{(s)}_\a$, with $s$ ranging over
all the sectors of the open string spectrum, are extracted from
the supersymmetric characters \be \chi^{(s)}_{_{SUSY}} = \sum_\a c_\a
\cZ^{(s)}_\a \ = \sum_\a c_\a {\theta_\a(0) \over \eta^3}
\cW^{(s)}_\a , \ee where $\cW^{(s)}_\a$ denotes the contribution
of the sector $s$ of the internal $\cN =2$ SCFT in a given spin structure $\a$
\cite{Dijkstra:2004ym,Dijkstra:2004cc}.

We would like to conclude with a comment on the supersymmetry
properties of our amplitudes. Some time ago
\cite{Berkovits:2001nv}, Berkovits and Vallilo have proposed
manifestly supersymmetric one-loop amplitudes for massless closed
string states based on the hybrid formalism
\cite{Berkovits:1996bf}. Deducing similar amplitudes for massless
open string states should be straightforward in the hybrid
formalism. However due to the factorization of the spacetime and
internal SCFT's in the hybrid formalism, in the absence of RR
fluxes, it is not clear to us how to reproduce the simple yet non
trivial internal structures, such as the functions $\cE_{\cN =1}$
and $\cJ_{\cN =1}$ that we have found in the NSR formalism. Other
manifestly supersymmetric formalisms \cite{Berkovits:2006bk,
Mafra:2005jh} may help clarifying this issue.  We leave it to
future work with eyes wide open to the possibilities of dealing
with RR fluxes \cite{Linch:2006ig} and the associated non-trivial
warping arising in flux compactifications \cite{Grana:2005jc}.

\section*{Acknowledgements}
We would like to thank Ya.~Stanev, E.~Trevigne, A.~Kumar,
M.~Serone, E.~Kiritsis, I. Klebanov for useful discussions. During
completion of this work M.B. was visiting the Galileo Galilei
Institute of Arcetri (FI), INFN is acknowledged for hospitality
and hopefully for support. M.B. would also like to thank the
organizers (C.~Angelatonj, E.~Dudas, T.~Gherghetta and A.~Pomarol)
and the participants to the workshop ``Beyond the Standard Model",
especially R.~Blumenhagen, for creating a stimulating environment
on the expectations for LHC\footnote{The most credited version
being now: {\bf L}'italia {\bf H}a vinto il {\bf C}ampionato del
mondo.}. This work was supported in part by INFN, by the
MIUR-COFIN contract 2003-023852, by the EU contracts
MRTN-CT-2004-503369 and MRTN-CT-2004-512194, by the INTAS contract
03-516346 and by the NATO grant PST.CLG.978785.
\newpage
\begin{appendix}

\section{Elliptic functions}

\subsection{Definitions}

Let $q=e^{2\pi i \tau}\,$ the Jacobi $\theta$-functions are
defined as gaussian sums \be \theta\left[^\a_\b\right](z|\tau)
=\sum_{n} q^{{1\over 2} (n +\a)^2} e^{2\pi i (z-\b)(n-\a)} \ , \ee
where $\a \ \b \in {\bf R}$.\\
Equivalently, for  particular values of characteristics, such as
$\a,\b = 0, 1/2$  they are  given also  in terms of infinite
product as follows \bea
\theta\left[^\frac{1}{2}_\frac{1}{2}\right](z|\tau) =
\theta_1(z|\tau) & = & 2 q^{1\over 8}\sin(\pi z)
\prod_{m=1}^{\infty}(1-q^m)(1-e^{2\pi i z}q^m)(1-e^{-2\pi i z}q^m)\nonumber\\
\theta\left[^\frac{1}{2}_0\right](z|\tau) = \theta_2(z|\tau) & = & 2
q^{1\over 8}\cos(\pi z)
\prod_{m=1}^{\infty}(1-q^m)(1+e^{2\pi i z}q^m)(1+e^{-2\pi i z}q^m)\nonumber\\
\theta\left[^0_0\right](z|\tau) = \theta_3(z|\tau) & = &
\prod_{m=1}^{\infty}(1-q^m)
(1+e^{2\pi i z}q^{m-\frac{1}{2}})(1+e^{-2\pi i z}q^{m-\frac{1}{2}})\nonumber\\
\theta\left[^0_\frac{1}{2}\right](z|\tau) = \theta_4(z|\tau) & = &
\prod_{m=1}^{\infty}(1-q^m) (1-e^{2\pi i
z}q^{m-\frac{1}{2}})(1-e^{-2\pi i z}q^{m-\frac{1}{2}}) \ . \eea

Dedekind $\eta$  function is defined as \be \eta(\tau) =
q^{1\over 24}\prod_{n=1}^{\infty} (1-q^n)  \ee and satisfies
$\th_1'(0) = 2\pi \et^3$.

Weierstrass $\cP$ function \be \cP(z) = - \de_z^2 \log \th_1(z) - 2
\et_1 \ , \ee where \be \et_1 = - 2\pi i {\de \over \de
\tau}\log\et = - {1\over 6} {\th_1'''(0) \over \th_1'(0)} \ , \ee
has a double pole at $z=0$ and is bi-periodic in $z$.

The free fermionic propagator in the even spin structures (Szego
kernel) \be \cS_\a(z) = {\th_\a(z0) \over \th_1(z)} { \th_1'(0)
\over \th_\a'(0)} \ee has a simple pole  at $z=0$ and satisfies
\be \cS_\a(z)^2 = \cP(z) - e_{\a-1} \ , \ee where \be e_{\a-1}=
4\pi i {\de \over \de \tau}\log{\th_\a(0)\over\et} \ee are also
related to $\cP(z)$ evaluated at the semiperiods \be e_1 =
\cP\left({1\over 2}\right) \ , \quad e_2 = \cP\left({1 + \tau\over
2}\right) \ , \quad e_3 = \cP\left({\tau\over 2}\right) \ . \ee

In the odd spin structure \be \cS_1(z) = -\de_z \cG(z) \ee is
biperiodic with a simple pole but not analytic.

The free bosonic propagator (biperiodic with logarithmic behaviour
at $z=0$) on the (covering) torus is given by \be \cG(z) =
-{\ap\over 2} \left[ \log\left\vert {\th_1(z) \over
\th_1'(0)}\right\vert^2 - {2\pi \over Im\tau} Imz^2 \right] \ .
\ee

\subsection{Pseudo-periodicity and zeroes}
Under lattice shifts of their first argument $z$, theta functions
transform according to  \bea &&\theta\left[^\a_\b\right](z+1|\tau)
= e^{2\pi i \a}
\theta\left[^\a_\b\right](z|\tau) \\
&&\theta\left[^\a_\b\right](z+\t|\tau) =  e^{-2\pi i (z+\b)-i\pi
\t} \theta\left[^\a_\b\right](z|\tau) \ . \eea

The location of their zeroes is given by\be
\theta\left[^\a_\b\right](z|\tau) = 0 \quad \leftrightarrow \quad
z_{n,m} = (\a - {1\over 2} +n ) \t + (\b - {1\over 2} + m ) \ .
\ee

\subsection{Modular Transformations}

Under T and S modular trasformations of their arguments theta
functions transform according to \bea
\theta\left[^\a_\b\right](z|\tau +1) & = & e^{-i\pi\a(\a-1)}\
\theta\left[^\a_{\b+\a-\frac{1}{2}}\right](z|\tau) \nonumber\\
\eta(\tau + 1) & = & e^{{i\pi\over12}}\,\eta(\tau) \nonumber\\
\theta\left[^\a_\b\right](\frac{z}{\tau}|-{1\over \tau} ) & = &
(-i\tau)^ {1\over2} \,e^{2i\pi\a\b +i\pi z^2/\tau}\
\theta\left[^\b_{-\a}\right](z|\tau)
\nonumber\\
\eta(-{1\over\tau}) & = & (-i\tau)^{1\over 2} \,\eta(\tau) \ .
\eea The modular transformation P, that connects the direct and
transverse channel of M\"obius strip amplitudes, is more involved.
It consists in a sequence of T and S transformations ($P=TST^2S$)
on the modular parameter $\tau_M = \frac{1}{2} + \frac{it}{2}$
\bea \theta\left[^\a_\b\right]({z\over i t}|\frac{1}{2} +
\frac{i}{2t}) & = & e^{-i\pi\a(\a-1) - 2\pi i (\a+\b-1/2)^2 +2\pi
z^2/t}\, \sqrt{-it}\; \theta\left[^{\a+2\b-2}_{1/2
-\a-\b}\right](z|\frac{1}{2} + \frac{it}{2})
\nonumber\\
\eta(\frac{1}{2} + \frac{i}{2t}) & = & e^{i\pi/4} \sqrt{-it}\;
\eta(\frac{1}{2} + \frac{it}{2}) \ . \eea

\subsection{Useful Identities}

Riemann identity for even spin structures reads \bea &&\sum_\a
c_\a \theta_\a(z_1)
\theta_\a(z_2)\theta_\a(z_3)\theta_\a(z_4) = \nn \\
&&\theta_1(z'_1) \theta_1(z'_2)\theta_1(z'_3)\theta_1(z'_4) -
\theta_1(z''_1) \theta_1(z''_2)\theta_1(z''_3)\theta_1(z''_4) \ ,
\eea where $z_i'$ and $z_i''$ are related to $z_i$ through \bea
&&z_1' = {1\over 2} (z_1 + z_2 +z_3 +z_4 ) \qquad z_2' = {1\over
2}
(z_1 + z_2 -z_3 -z_4 ) \nn \\
&&z_3' = {1\over 2} (z_1 - z_2 +z_3 -z_4 ) \qquad z_4' = {1\over
2} (z_1 - z_2 -z_3 + z_4 ) \eea and \bea &&z_1'' = {1\over 2}
(-z_1 + z_2 +z_3 +z_4 ) \qquad z_2'' = {1\over 2}
(z_1 - z_2 +z_3 +z_4 ) \nn \\
&&z_3'' = {1\over 2} (z_1 + z_2 -z_3 +z_4 ) \qquad z_4'' = {1\over
2} (z_1 + z_2 +z_3 - z_4 ) \ . \eea

\subsection{Series Expansions}

Series expansion in powers of $q$ yield

\bea \de_z \log\theta_1(z,q) &=& \pi \cot(\pi
z) + 4\pi \sum_n { q^n \over 1-q^n} \sin(2\pi n z) \nn \\
&=& \pi \coth(\pi z) + 4\pi \sum_{n, d_n | n} q^n \sin(2\pi d_n z)
\eea and
 \bea
\de^2_z \log\theta_1(z,q) &=& -{\pi^2\over \sin(\pi z)^2} + 8\pi^2
\sum_n { n q^n \over 1-q^n} \cos(2\pi n z) \nn \\
&=&   -{\pi^2\over \sin(\pi z)^2} + 8\pi^2 \sum_{n, d_n | n} q^n
d_n \cos(2\pi  d_n z) \ , \eea where, using $\de_\tau q = 2\pi i
q$, \be \eta_1 \equiv -2\pi i \de_\tau \log\eta = {\pi^2\over 6} -
4\pi^2\sum_n { n q^{n} \over 1-q^n} = {\pi^2\over 6} - 4\pi^2
\sum_{n, d_n | n} q^n d_n \ee so that \be\cP(z) \equiv -\de_z^2
\log\theta_1(z,q) - 2\eta_1 = {\pi^2\over \sin(\pi z)^2} -
{\pi^2\over 3} + 8\pi^2 \sum_{n, d_n | n} q^n d_n [1 - \cos(2\pi
d_n z)] \ . \ee

Moreover \be 4\pi i \de_\tau \log\theta_1(z,q) = -{\pi^2} + 8\pi^2
\sum_{n, d_n | n} q^n {n \over d_n}[1 + 2 \cos(2\pi d_n z)] \ee
and \bea  \left({\theta_1'(z,q)\over \theta_1(z,q)}\right)^2 &=&
\de^2_z \log\theta_1(z,q)  - {\theta_1''(z,q)\over \theta_1(z,q)}
= \de^2_z \log\theta_1(z,q) - 4\pi i \de_\tau
 \log \theta_1(z,q) \nn \\
&=& - \pi^2 cot(\pi z)^2  - 8\pi^2 \sum_{n, d_n | n} q^n \left[
\left({2 n \over d_n}- d_n\right) \cos(2\pi d_n z) + { n \over
d_n} \right] \ . \nn \\
\eea

For points on different boundaries in the transverse channel
$\theta_1$ gets effectively replaced by $\theta_4$ for which \bea
\de_z \log\theta_4(z,q) &=& 4\pi \sum_n { q^{n/2} \over 1-q^n}
\sin(2\pi n z) \nn \\
&=& 4\pi \sum_{n, d_n | n} q^{n/2} {1 - (-)^{n/d_n} \over 2}
\sin(2\pi d_n z) \eea and \bea \de^2_z \log\theta_4(z,q) &=&
8\pi^2 \sum_n { n q^{n/2}
\over 1-q^n} \cos(2\pi n z) \nn \\
&=& 8\pi^2 \sum_{n, d_n | n} q^{n/2} {1 - (-)^{n/d_n} \over 2} d_n
\cos(2\pi d_n z) \ . \eea

Moreover \bea 4\pi i \de_\tau \log\theta_4(z,q) &=&  8\pi^2 \sum_n
\left[ { n q
\over 1-q^n} + {2 q^n \cos(2\pi n z) \over (1-q^{2n})^2} \right] \\
&=& 8\pi^2  \sum_{n, d_n | n} \left[ d_n q^{n} + {1 -
(-)^{n/d_n}\over 2} {n\over d_n} q^{n/2} \cos(2\pi d_n z)\right]
\nn  \eea so that
 \bea \left({\theta_4'(z,q)\over \theta_4(z,q)}
\right)^2 &=& \de^2_z \log\theta_4(z,q)  -  {\theta_4''(z,q)\over
\theta_4(z,q)} = \de^2_z \log\theta_4(z,q)  -  4\pi i {\de_\tau
\theta_4(z,q)
\over \theta_4(z,q)} \\
&=& - 8\pi^2  \sum_{n, d_n | n} \left[ d_n q^{n} + {1 -
(-)^{n/d_n}\over 2} \left({n\over d_n} - d_n\right) q^{n/2}
\cos(2\pi d_n z)\right] \nn \ . \eea

\end{appendix}

\end{document}